\documentclass{article}

\usepackage{graphicx}
\usepackage{cite}
\usepackage{epsfig}
\usepackage{amsfonts}
 \usepackage{amssymb}
\usepackage{amsmath}
\usepackage{slashed}
\usepackage{gensymb}

\usepackage{hyperref}
\usepackage{graphicx}       
\usepackage{dcolumn}        
\usepackage{bm}            		 
\usepackage{amssymb}
\usepackage{amstext}
\usepackage{tensor}
\usepackage{color} 
\usepackage{transparent}

\date{\today}

\newcommand{\insertplot}[5]{\begin{figure}
 \hfill\hbox to 0.05in{\vbox to #5in{\vfill
 \inputplot{#1}{#4}{#5}}\hfill}
 \hfill\vspace{-.1in}
 \caption{#2}\label{#3}
 \end{figure}}
 \newcommand{\inputplot}[3]{
 \special{ps: plotfile #1}
}
\newcounter{fig}

\newcommand{\ee}{\end{equation}}
\newcommand{\eea}{\end{eqnarray}}
\newcommand{\bea}{\begin{eqnarray}}

\newcommand{\beq}{\begin{equation}}
\newcommand{\eeq}{\end{equation}}

\newcommand{\ze}{\kern 0.05em}

\begin{document}
\begin{center}

{\LARGE \bf The imitation game: \\ {Proca stars that  can mimic the Schwarzschild shadow}}
\\
\vspace{0.8cm}
{{\bf Carlos A. R. Herdeiro$^{\ddagger}$,  
Alexandre M. Pombo$^{\ddagger}$,
Eugen Radu$^{\ddagger}${, \\ Pedro V. P. Cunha$^{\ddagger}$}   and Nicolas Sanchis-Gual$^{\ddagger,\dagger}$
}
\vspace{0.3cm}
\\
$^{\ddagger }${\small Departamento de Matem\'atica da Universidade de Aveiro and } \\ {\small  Centre for Research and Development  in Mathematics and Applications (CIDMA),} \\ {\small    Campus de Santiago, 3810-183 Aveiro, Portugal}\\
$^{\dagger }${\small 
Departamento  de  F\'\i sica,  Instituto  Superior  T\'ecnico  -  IST,} \\ {\small  Universidade  de  Lisboa  -  UL,  Avenida  Rovisco  Pais  1,  1049-001,  Portugal}\\
}
\vspace{0.3cm}
\end{center}

\date{April 2020}

\begin{abstract}   
Can a \textit{dynamically robust} bosonic star (BS) produce an (effective) shadow that mimics that of a black hole (BH)? We focus on  models of spherical BSs with free  scalar or vector fields, as well as with polynomial or axionic self-interacting fields. The  BH shadow is linked to the existence of light rings  (LRs). For free bosonic fields, yielding  \textit{mini}-BSs, it is known that these stars can become ultra-compact - $i.e.$, possess LRs - but only for perturbatively unstable solutions. We show this remains the case even when different self-interactions are considered. 
However, an effective shadow can arise in a different way: if BSs reproduce the  existence of an innermost stable circular orbit (ISCO) for timelike geodesics (located at  $r_{\rm  ISCO}=6M$ for a Schwarzschild BH of mass $M$), the accretion flow morphology around BHs is mimicked and an effective shadow arises in an astrophysical environment. Even though spherical BSs  may accommodate stable timelike circular orbits all the way down to their  centre, we show the angular velocity $\Omega$ along  such orbits may have a maximum away from the origin, at $R_{\Omega}$; this scale was recently observed to mimic the BH’s ISCO in some scenarios of accretion flow. Then: $(i)$ for free scalar fields or with quartic self-interactions, $R_{\Omega}\neq 0$ only  for  perturbatively unstable BSs; $(ii)$ for higher scalar self-interactions, $e.g.$ axionic, $R_{\Omega}\neq 0$  is  possible for perturbatively stable BSs, but no solution with $R_{\Omega}=6M$ was  found in the parameter space explored; $(iii)$  but for free vector fields, yielding Proca stars,  perturbatively stable solutions with  $R_{\Omega}\neq 0$ exist, and indeed  $R_{\Omega}=6M$ for a particular solution. {Thus, dynamically robust spherical  Proca  stars succeed in the imitation game: they can  mimic the shadow of a (near-)equilibrium Schwarzschild BH with the same $M$, in an astrophysical environment, despite the  absence of a LR, at least under some observation conditions, as we confirm by explicitly comparing the lensing of such Proca stars and Schwarzschild BHs}.  
\end{abstract}

\tableofcontents

%
 \section{Introduction}\label{S1}
%
Bosonic stars (BSs) are speculative macroscopic Bose-Einstein condensates. They may be described as everywhere regular lumps ($i.e.$~self-gravitating solitons)  of yet undetected ultralight scalar~\cite{kaup1968klein,ruffini1969systems} or vector~\cite{brito2016proca} bosonic fields - see also, e.g.~\cite{colpi1986boson,lynn1989q,Schunck:1996he,Yoshida:1997qf,astefanesei2003boson,schunck2003general,Liebling:2012fv,Grandclement:2014msa,herdeiro2017asymptotically,Alcubierre:2018ahf,Herdeiro:2019mbz,guerra2019axion,
Delgado:2020udb,minamitsuji2018vector,Herdeiro:2020kvf,Herdeiro:2020jzx} for some generalizations and reviews. 
Such hypothetical ultralight bosons could be part (or the whole) of the dark matter budget in  the Universe~\cite{CruzOsorio:2010qs,Suarez:2013iw,Hui:2016ltb}. In particular,  compact and dynamically robust BSs  occur in a variety of different models~\cite{Liebling:2012fv}, thus being  interesting for a diversity of theoretical and phenomenological strong gravity  studies - see e.g.~\cite{kusmartsev1991stability,gleiser1988stability,gleiser1989gravitational,hawley2000boson,Palenzuela:2007dm,guzman2009three,LoraClavijo:2010xc,
Eilers:2013lla,Cunha:2015yba,cunha2016chaotic,Cao:2016zbh,Shen:2016acv,
vincent2016imaging,Franchini:2016yvq,cunha2017lensing,grandclement2017light,sanchis2017numerical,cardoso2017tests,Palenzuela:2017kcg,
grould2017comparing,Bezares:2017mzk,glampedakis2018well,Croon:2018ftb,olivares2018tell,Johnson-McDaniel:2018uvs,CalderonBustillo:2020srq,Siemonsen:2020hcg}.  In fact, BSs have been suggested as possible black hole (BH) mimickers~\cite{Cardoso:2019rvt}. The purpose of this work is to assess this possibility in what concerns  the BH shadow~\cite{Falcke:1999pj,Cunha:2018acu}, for equilibrium (or near-equilibrium) BSs.

An important feature of the paradigmatic BH model of general relativity, the Kerr BH~\cite{Kerr:1963ud}, is the existence of bound photon orbits (see $e.g.$ \cite{Bardeen:1972fi}), which, in their simplest guise are planar light rings (LRs). Furthermore, LRs have been show to be a  generic feature of asymptotically flat stationary BHs, even beyond vacuum or beyond Einstein's theory~\cite{Cunha:2020azh}. The existence of LRs around BHs impacts on important  strong  gravity features, such as (the initial part of)  the  ringdown~\cite{Cardoso:2016rao} and the BH shadow~\cite{Cunha:2017eoe}. Thus, it has been generically assumed that in order to mimic these features, BSs should possess LRs. For the simplest models of BSs, where the bosonic field has no self-interactions, leading to the so-called \textit{mini}-BS, such ultracompact solutions indeed exist, but only in special regions of the parameter  space, where the BSs  have been shown to be  unstable~\cite{cunha2017lensing}. One may wonder,  nonetheless, whether in other models, possessing self-interactions, ultracompact BSs could  arise for perturbatively stable BSs. As we show in this paper, however, this is not the case for several examples of self-interactions. 

Moreover, it has been  shown that topologically trivial spacetime configurations, such as BSs,  develop not one, but two LRs, when they become ultracompact~\cite{cunha2017light}. And, if the bosonic matter obeys the null energy condition (which  is the case for the standard bosonic fields considered) one of the LRs is stable. Such stable LRs  have been argued to  lead to a non-perturbative instability~\cite{Keir:2014oka,Cardoso:2014sna,Benomio:2018ivy}. Little is known about the timescales of this putative instability, which, therefore, is  not an unsurmountable obstacle \textit{per se} for ultracompact  BSs to be dynamically robust. Nonetheless, together  with the inability to have perturbatively stable ultracompact BSs, this feature casts an additional shadow of doubt on the dynamical  viability of ultracompact BSs.

There is, however, a different possibility allowing a BS without LRs  to mimic the appearence of a BH, when lit by a surrounding accretion flow. If the source of  light in the vicinity of the BS has the same morphology as it would have around a BH, the lensing of light,  and in particular a  similar central depression of  the emission (the  shadow~\cite{Falcke:1999pj}) would also be present~\cite{vincent2016imaging,grould2017comparing}. The key feature here is the cut-off in the emission due to the disk's inner edge, which is determined by the innermost stable circular orbit (ISCO)  of the BH, for timelike geodesics. For a Schwarzschild BH of mass $M$, the ISCO is located at the areal  radius $r=6M$. For spherical BSs there is no ISCO, so that one could think that  the disk, and the emission,  continues all the way  to the centre: hence no shadow should be produced. In a recent work~\cite{olivares2018tell}, however, general relativistic magnetohydrodynamic  simulations were performed  in static (scalar) BSs backgrounds, including general relativistic radiative transfer, observing qualitative similarities with BH spacetimes, in particular the central emission depression, in other words, an effective shadow. The key new
feature observed in~\cite{olivares2018tell} is that despite the existence of stable timelike circular orbits in the BSs spacetime all the way up to the   centre,  the angular velocity of the orbits, $\Omega$, attains a  maximum at some areal radius ($R_{\Omega}$).  This scale is observed to  determine  the inner edge of the accretion disk in the simulations in~\cite{olivares2018tell}, under the assumptions therein, namely  that the loss of angular momentum of the orbiting matter is driven  by  the  magneto-rotational  instability~\cite{Balbus:1991ay} and  that the radiation relevant for BH shadow observations is mostly due to synchroton emission. 

Despite the qualitative similarity,  $i.e.$ the possibility of obtaining an effective shadow in a BS spacetime and  in a realistic astrophysical environment (despite the absence of LRs  or ISCO), the results in~\cite{olivares2018tell} raise two issues. Firstly, they show a quantitative difference between  the BS and BH shadow for the cases analysed. For the same total mass,  the two have a distinguishable size, even by current observations. Secondly, and most importantly, the BSs in  the analysis in~\cite{olivares2018tell} that  display the new scale $R_{\Omega}$ are perturbatively unstable. Hence, one may wonder whether there are models in which spherical BSs can have a degenerate (effective) shadow with a comparable BH ($i.e.$ with the same  ADM  mass) in the perturbatively stable region. As we shall see below:  $(i)$ self-interactions of scalar bosonic fields can indeed yield perturbatively stable BS with the new scale $R_{\Omega}$; however  in  this case we could  not get solutions with $R_{\Omega}=6M$; $(ii)$ but for  vector BSs [aka \textit{Proca Stars} (PSs)], even without self-interactions, there are perturbatively stable stars with the new scale $R_{\Omega}$ and we find a particular solution with $R_{\Omega}=6M$. Thus, models in which dynamically robust spherical BSs can have a degenerate (effective) shadow with a comparable Schwarzschild BH \textit{do exist}. {We confirm this possibility by explicitly studying  the lensing of the aformentioned particular PS with  $R_{\Omega}=6M$, lit by an  accretion disk with its inner edge at this  special radius. The analysis, however,  clarifies that the degenerate shadow only occurs for some  observation angles.}

This paper is organised as follows. In Sec.~\ref{S2} we introduce the generic model for BSs and the field equations for both the scalar and vector  cases. We compute the required asymptotic expansions at the center of the star and at infinity. These are used to numerically compute the BSs solutions  of which  we display the  domain of  existence in the following sections. In Sec.~\ref{S3} we derive the LR and timelike circular orbits' (TCOs) relations. These are then investigated for the scalar BSs cases in Sec.~\ref{S4} and for the PSs in Sec.~\ref{S5}. We conclude with a summary of our main result and a discussion.

\bigskip

%
	\section{Models}\label{S2} 
%
	The Einstein-matter action, where the matter part describes a spin $s=0,1$ classical field minimally coupled to Einstein's gravity, reads
	\begin{equation}
	 \mathcal{S}=\int d^4 x \sqrt{-g} \left[ \frac{R}{16 \pi G}+\mathcal{L}_s \right]\ ,
	\end{equation}
where $R$ is the Ricci scalar of the spacetime represented by the metric $g_{\alpha \beta}$ with metric determinant $g$, $G$ is Newton's constant and 
	the matter Lagrangians for the spin 0 and  spin 1 fields are, respectively:
	\begin{equation}
	\mathcal{L}_0 = 
	-\frac{1}{2}g^{\alpha \beta} \big(\bar{\Phi} _{,\alpha} \Phi _{,\beta}+\bar{\Phi} _{,\beta}\Phi _{,\alpha} \big) - U_i(|\Phi |^2)\ , \qquad \qquad \mathcal{L}_1= -\frac{1}{4} F_{\alpha \beta}\bar{F} ^{\alpha \beta}-V (\textbf{A}^2)\ .
	\end{equation}
	 The massive complex scalar field, $\Phi$, with mass $\mu _S$, has a potential term $U_i(|\Phi |^2)$; the massive complex vector field, with mass $\mu _P$, has a $4$-potential $A^\alpha$ and is under a potential  $V (\textbf{A}^2)$. We have used the notation $\textbf{A}^2\equiv A_\alpha \bar{A}^\alpha$ and an overbar denotes complex conjugation. 

Variation of the action with respect to the metric and matter fields leads to the following two sets of field equations, in the scalar and vector case, respectively (setting $ G=1$)
	\begin{align}
	 & G_{\alpha \beta} = { 4\pi} \Big[\bar{\Phi} _{,\alpha} \Phi _{,\beta}+\bar{\Phi} _{,\beta}\Phi _{,\alpha}-g_{\alpha \beta} \mathcal{L}_0 \Big]\ , \qquad \Box \Phi = \hat{U_i}\ \Phi \ ,\\
	 & G_{\alpha \beta} ={4\pi  }\left[\frac{1}{2} \big(F_{\alpha \delta} \bar{F}_{\beta \gamma} +\bar{F}_{\alpha \delta} F_{\beta \gamma} \big) g^{\delta \gamma}+\hat{V}\big( A_\alpha \bar{A}_\beta +\bar{A}_\alpha A_\beta - g_{\alpha \beta}\mathcal{L}_1 \big)\right] \ , \qquad\frac{1}{2} \nabla_\alpha F^{\alpha \beta} = \hat{V} A^\beta\ ,
	\end{align}
	with $G_{\alpha \beta}$ the Einstein's tensor, $\Box$  $(\nabla)$ the covariant d'Alembertian (derivative) operator, $\hat{U_i}\equiv d U_i/d |\Phi |^2$ and $\hat{V}\equiv dV /d \textbf{A}^2$.

For the metric ansatz, we use a spherically symmetric solution with two unknown functions,
	\begin{equation}\label{E24}
	ds^2 = -N(r) \sigma ( r) ^2 dt^2+\frac{dr^2}{N(r)}+r^2\big(d\theta ^2 + \sin ^2 \theta d\phi ^2 \big)\ ,
	\end{equation}
	with $N(r)\equiv 1-\frac{2 m(r)}{r}$, $m(r)$ the Misner-Sharp mass function~\cite{Misner:1964je} and $\sigma (r)$ the second unknown metric function. The matter field ansatz reads, for the scalar and vector cases, respectively:
	\begin{equation}\label{E25}
	\Phi (r,t) = \varphi (r) e^{-i \omega t}\ ,\qquad \qquad A=\big[ f(r) dt + \textit{i} g(r) dr \big] e^{-i\omega t}\ ,
	\end{equation}
	where $\varphi (r) $ is  the scalar field amplitude and $f(r)$ and $g(r)$ are two real potentials that define the Proca ansatz. In both cases $\omega $ is the field's frequency. In the Proca case, the field equation implies, for a Ricci-flat space,
	the Lorenz condition $\nabla _\alpha (\hat{V}A^\alpha) =0$, which is a dynamical condition, rather than a gauge choice.
	
	Both matter models possess a $\textbf{U}(1)$ global symmetry, under a global phase transformation: $\Phi \rightarrow \Phi e^{i a}$ and $A \rightarrow A e^{i a}$, where $a$ is a constant. This symmetry leads to a conserved Noether charge, $Q$, from the spatial integration of the time component of the conserved Noether current ($Q =\int _\Omega j_s ^t$), with
	\begin{equation}\label{W26}
	j_0 ^\alpha = -\textit{i}\ \big[ \bar{\Phi} \partial ^\alpha \Phi -\Phi \partial ^\alpha \bar{\Phi} \big]\ ,\qquad \qquad j_1 ^\alpha = \frac{i}{2}\big[\bar{F} ^{\alpha \beta} A_\beta -F^{\alpha \beta}\bar{A}_\beta\big]\ ,
	\end{equation}
where the subscripts in $j$ refer to the model $s=0,1$. A similar notation is used below for other quantities.
	For the vector model, we will focus on the following self-interactions  potential:
	\begin{equation}
	V=\frac{\mu _P ^2}{2} \textbf{A}^2+\frac{\lambda _P}{4} \textbf{A}^4\ .
\label{vp}
	\end{equation}	 
	This model encompasses the mini-PSs solutions~\cite{brito2016proca} when the self-interaction coupling vanishes, $\lambda _P =0$. 

With the above setup one obtains a system of coupled Einstein-matter ordinary differential equations. For each of the two models ($s$=0, 1)  there are two ``essential" Einstein equations:
	\begin{eqnarray}
	\label{E28}
&& s=0:~~
	m' = {4\pi} r^2 \left[N (\varphi ') ^{2}+\frac{\omega ^2 \varphi ^2}{N \sigma ^2}+U_i \right]\ , ~~
	\sigma ' = {8\pi} \sigma r \Big[ (\varphi ') ^{2}+\frac{\omega ^2 \varphi ^{2}}{N ^2 \sigma ^2}\Big]\ 
	          \\
&& s=1:~~
	m' = {4\pi}r^2 
	\left[ \frac{(f'-\omega g)^2}{2\sigma ^2} +
	\left(\mu _P ^2 -\frac{3}{2}\lambda_P \textbf{A}^2\right)\frac{f^2}{2 N \sigma^2}+\frac{V}{\textbf{A}^2}N\right] \ , ~~
	\sigma ' = {8\pi} r \sigma \hat{V} \left[ g^2+ \frac{f^2}{N^2 \sigma ^2}\right].\quad  \quad \quad 
	\label{E29}	
	\end{eqnarray}
	To  close the system, the equations for the matter field functions are
	\begin{eqnarray}\label{E210}
	\varphi '' &=& -\frac{2 \varphi '}{r}-\frac{N' \varphi '}{N}-\frac{\sigma ' \varphi '}{\sigma}-\frac{\omega ^2 \varphi}{N^2 \sigma ^2}+{\frac{\hat{U_i}}{N}}\ \varphi\ ,\\
	f' &=& \omega g-2\frac{g \sigma ^2 N}{\omega} \hat{V}\ , \qquad \qquad \frac{d}{dr}\left[\frac{r^2\big(\omega g - f'\big)}{\sigma}\right]+\frac{2 r^2 f}{N \sigma}\hat{V} =0\ .  \label{E211}
	\end{eqnarray}
%
	\subsection{Asymptotic expansions and physical relations}\label{S22}
%
In order to integrate   the field equations  \eqref{E28}-\eqref{E211}  one must  consider the asymptotic expansions. At the origin, the field equations can be approximated by a power series expansion in $r$ that guarantees $m (0)=0,\ \sigma (0) =s _0 ,\ \varphi (0)=\varphi _0 , \ f(0)=f_0$ and $g(0)=0$
\begin{eqnarray}
\nonumber
&& s=0:~~
	  m  \approx {\frac{4\pi}{3}} \frac{U_i s _0 ^2+\omega ^2 \varphi _0 ^2}{s _0} r^3 + ...\ ,
		\quad  
\sigma  \approx  s _ 0 + {4\pi}\frac{\omega ^2 \varphi _0 ^2}{s _0} r^2 + ...\ ,
\quad 	 
\varphi \approx  \varphi _0 +\frac{\varphi _0}{6} \left( \hat{U_i}-\frac{\omega ^2}{s _ 0 ^2}\right) r^2 +... \ , 
\quad 	
		\\
		\label{E212}
&& s=1:	~~
 m \approx  {\frac{4\pi}{6}}\frac{f_0 ^2 \mu _P ^2 s _0 ^2-\frac{3 f_0 ^4 \lambda _P }{2}}{ s _0 ^4} r^3 + ...\ , \qquad 
\sigma  \approx  s _0 + 2\pi\frac{ f_0 ^2 \mu _P ^2 s _0 ^2+ f_0 ^4 \lambda _P}{ s _0 ^3} r^2 + ...\ , 
\\
\nonumber
&&~~~~~~~~f \approx f_0 -\frac{f_0 \left(f_0 ^2 \lambda _P -\mu _P ^2 s _0 ^2+\omega ^2\right)}{6 s _0 ^2} r^2 +... \ ,\qquad g \approx -\frac{f_0 \omega }{3s _0 ^2} r+...\label{E215}\ .\quad \quad 
	\end{eqnarray}

	At infinity, we impose asymptotic flatness and a finite ADM mass $M$: $m(\infty ) =M,\ \sigma (\infty) = 1,$  and $ \varphi (\infty) =0=f(\infty)=g(\infty)$. The values of $s _0$ and $M$ are fixed by the numerics, while $\sigma (\infty)$ fixes  the following scaling symmetry of  the system of equations: $\{ \sigma, \omega, f_0 \} \rightarrow \zeta \{ \sigma , \omega, f_0 \}$, with $\zeta >0$. An additional reescaling symmetry, moreover,  allows both mass's $\mu _P $ and $\mu _S $ to be set to unity
	$\big( \mu _P  = \mu _S =1 \big)$.

	The set of coupled ODE's are numerically integrated by means of a $6^{\rm th}$ order adaptative step Runge-Kutta method, with a local error of $10^{-15}$. The boundary conditions are enforced using a shooting strategy, with a tolerance of $10^{-9}$ for the spatial asymptotic (at infinity)  scalar/Proca field decay value, while $m(r)\rightarrow M$ and $\sigma \rightarrow 1$. 
		
	Due to the lack of a surface, BSs do not have a well-defined radius. For the ``radius" of BSs we will consider the areal radius of  a spherical surface within which $99\% $ of all the mass is included; this radius is denoted $R_{99}$. The latter defines the BS compactness: $\mathcal{C}\equiv 2M_{99}/R_{99}$. This  compactness is always smaller than unity, becoming unity for BHs.

	To test the numerical solutions, we have considered the so-called virial identities, a set  of identities obtained from a Derrick-type scaling argument~\cite{Derrick:1964ww}, and which are independent from the equations of motion. These read for the $s=0,1$ cases, respectively,
	\begin{eqnarray}
&& s=0:~~~\int _0 ^{\infty}dr\ r^2 \sigma \left[ \frac{\omega ^2 \varphi ^2}{N \sigma ^2}\left(3-\frac{2 m}{rN}\right)-\varphi^{' 2}-3U_i\right]=0\ ,\nonumber\\
&& s=1:~~~\int _0 ^\infty dr\ \frac{ r^2 }{2 N^3 \sigma ^3} \Bigg\{ f^4 (2-5 N) \lambda _P+2 f^2 N \sigma ^2 \Big(N \left(3 g^2 N \lambda _P +4\right)-1\Big)+ N^3 \sigma ^2 \Big[ 2 f' \left(f'-4 g \omega \right)\Big.\Bigg.\nonumber\\
&& \qquad\qquad\qquad\qquad \quad \qquad\Big.\Bigg.+g^2 \left(6 \omega ^2-\sigma ^2 \left(g^2 N (N+2) \lambda _P +4 N+2\right)\right)\Big]\Bigg\} =0\ .\nonumber
	\end{eqnarray}

	The numerical accuracy can also be tested by the ADM mass expressions computed as a volume integral, which  can be compared to the value of $M$ computed from the mass function at infinity. The volume integrals read
	\begin{eqnarray}
	\nonumber
&& s=0:~~~M= \int _0 ^\infty dr\ r^2 \sigma \left[\frac{4\omega ^2 \varphi ^2}{N\sigma ^2}-2 U_i\right]\ , 
\\
	\nonumber
&& s=1:~~~M=  \int _0 ^\infty dr\frac{ r^2}{4 N^2 \sigma ^4} 
                       \left[
-3 f^4 \lambda _P +N \sigma ^2 
\left ( 
2 f^2  \left(g^2 N \lambda _P+1\right)+2 N f^{' 2}+ N g^2 \left(g^2 N \lambda _P +2\right)-2 \omega g 
\right )
                       \right] .
	\end{eqnarray}
%
	\section{Light rings (LRs) and timelike circular orbits (TCOs)}\label{S3}
%
Let us now consider the basic equations to compute both LRs and TCOs  in BSs spacetimes.
	
	The radial geodesic equation for a particle around a BS, described by the metric \eqref{E24}, is
	\begin{equation}\label{E316}
	\dot{r}^2 =\frac{E^2}{\sigma ^2}-\frac{l ^2 N}{r^2}+k N\ , 
	\end{equation}
where $E,l$ represent the particle's energy and angular momentum and  the dot represents the derivative with respect to an affine parameter. For null (timelike) geodesics $k=0$ $(k=-1)$.

Let us first consider null geodesics $(k=0)$.	For a LR, $\dot{r}=0$, which relates $E$ and $l$, $E=\frac{l \sqrt{N} \sigma}{r}$. The LR is circular, which additionally imposes $\ddot{r}=0$. This gives the condition for the presence of a LR
	\begin{equation}\label{E318}
	 -r \sigma \left(\frac{-2 m'}{r}+\frac{2m}{r^2}\right)+2\left( 1-\frac{2m}{r}\right) \big( \sigma -r \sigma ' \big) =0\ .
	\end{equation}
	As shown in \cite{cunha2017light}, BSs' LRs always come in pairs -- one stable and one unstable -- corresponding to the two roots of \eqref{E318}. Here, we wish to find the first BS solution  containing a LR; in other words, the first ultracompact BS. Let $\omega _{LR}$ be the frequency of the first ultracompact BS, as we move along the  (one-dimensional) domain  of existence, starting from the Newtonian limit ($cf.$ Secs.~\ref{S4} and~\ref{S5} below). The latter corresponds to a BS solution with two degenerate LRs~\cite{cunha2017light}.

Let us now turn to TCOs ($k=-1$). They are described by the tangential $4-$velocity $u^\nu = \big( u^t, 0,0,u^\phi \big)$ with the normalization condition $u^2=u^\nu u_\nu =-1$. The angular velocity $\Omega$ along these orbits is
	\begin{equation}\label{E324}
	\Omega = \frac{u^\phi}{u^t} = \sqrt{\frac{\sigma}{2r}} \sqrt{\sigma N'+2 N \sigma '}\ .
	\end{equation}

	As argued in~\cite{olivares2018tell},  an accretion disk may have an inner edge even around BSs without an ISCO. This occurs if the angular velocity along TCOs attains a maximum at some radial distance. The  corresponding areal radius is denoted $R_\Omega$.  	This is   computed by monitoring the angular frequency $\Omega$ at all radial points to obtain its maximum.

	In the next sections we shall study the first ultracompact BS and the existence of $R_\Omega$ for several different models. As an accuracy estimate, for all the computed BSs solutions, the virial identity and mass relations are obeyed within a factor of $~10^{-9}$.
%

%

%
	\section{Scalar BSs}\label{S4}
%
Many scalar BS models have been considered over the years - see $e.g.$~\cite{schunck2003general}. Here, we shall divide our analysis into two different cases, depending on the choice of the self-interactions potential.

The first case considers a polynomial self-interaction of the type:
	\begin{equation}
	 U_{\rm poly}=\mu_S ^2 \Phi ^2 + \lambda\Phi ^4+ \gamma \Phi ^6\ ,
\label{poly}
	\end{equation}
	where $\gamma$ ($\lambda$) is a coupling controling the self-interaction of sixth (fourth) order. Thus, the potential is determined by two parameters, since we have already established that the mass can be set to unity, $\mu_S =1$.

There are three sub-cases of interest.   The most generic case occurs for $\gamma\neq 0 \neq \lambda$.
    The latter BSs are known as $Q$-Stars~\cite{lynn1989q}, since they are self-gravitating generalisations of the flat spacetime $Q$-balls~\cite{coleman1985q}, with the constants $\mu_S,\lambda$
and $\gamma$ subject to some conditions.
 For $\gamma=0$ and $\lambda>0$, one obtains the Colpi-Shapiro-Wassermann quartic scalar BSs~\cite{colpi1986boson}. For $\gamma=\lambda=0$ one recovers the scalar mini-BSs~\cite{kaup1968klein,ruffini1969systems}.

The second case considers a non-polynomial, axion-type potential~\cite{guerra2019axion,Delgado:2020udb}:
	\begin{equation}
	 U_{\rm axion}=\frac{2\mu_S ^2 f_\alpha ^2}{\hbar B}\left[ 1-\sqrt{1-4B\sin ^2 \left( \frac{\Phi \sqrt{\hbar}}{2 f_\alpha}\right) }\ \right]\ ,
\label{axionp}
	\end{equation}
	where $f_\alpha$ is the coupling strength and $B=\frac{z}{1+z^2}\approx 0.22$ with $z\equiv \frac{m_\mu}{m_d}\approx 0.48$ the mass ratio of the up/down quark. The second term in the potential is the standard QCD axion potential to which is added a constant term to ensure $U_{\rm  axion} (0)=0$ and hence asymptotical flatness. 
	
	The potential is characterized by two parameters: $f_\alpha $ and $\mu_S $. By expanding $U_{\rm axion} (\Phi ) $ around the minimum $\Phi =0$
	\begin{equation}\label{E324n}
	U_{\rm axion} (\Phi ) \approx \mu_S ^2 \Phi ^2 -\left(\frac{3B-1}{12}\right) \frac{\hbar \mu _S ^2}{f_\alpha ^2}\Phi ^4 + \mathcal{O}(\Phi ^6) \ ,
\end{equation}		
	we identify the axion-like particle mass and a quartic self-interaction coupling ($\lambda /2$), respectively,
	\begin{equation}
	m_{\rm axion} = \mu_S \hbar \ ,\qquad \qquad \left( \frac{3B-1}{12}\right) \frac{\hbar \mu_S ^2}{f_\alpha ^2} = \frac{\lambda}{2} \ .
	\end{equation}
	A decrease in the coupling strength $f_\alpha$ implies a decrease in the width of the potential; this is equivalent to an increase in the self-interaction, $f_\alpha \propto 1/\sqrt{\lambda}$.  Thus,  the mini-BS model is recovered as $f_\alpha\rightarrow \infty$.

%

	\subsection{The polynomial self-interaction:  $\gamma=0$}\label{S51}
%

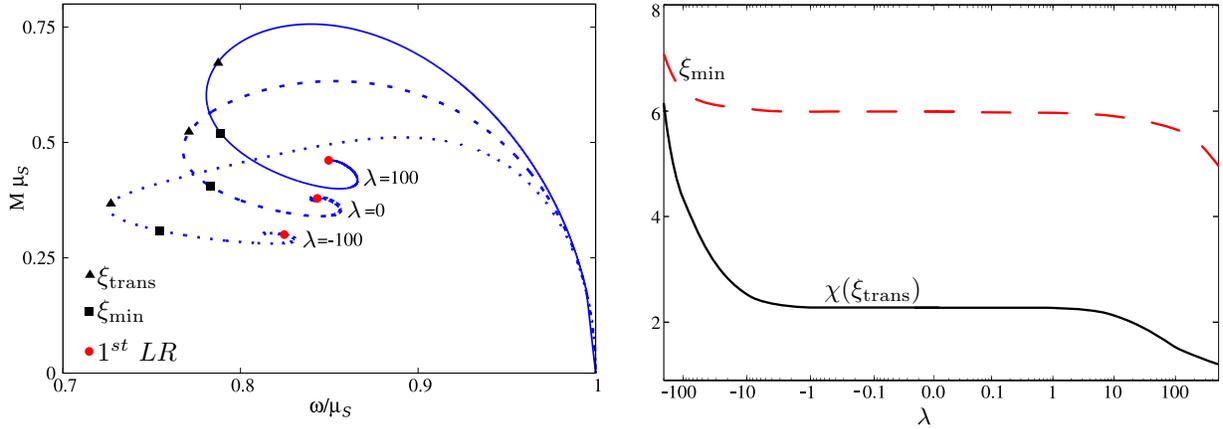
\begin{figure}[h]
		 \centering
		 	\begin{picture}(0,0)
			 \put(41.5,24){$1^{st}\ LR$}
			 \put(41.5,40){$\xi_{\rm min}$}
			 \put(41.5,53){$\xi _{\rm trans}$}
			 \put(142,90){\small{$\lambda $}}
			 \put(136,78){\small{$\lambda $}}
			 \put(119,67){\small{$\lambda $}}
			\end{picture}
		 \includegraphics[scale=0.65]{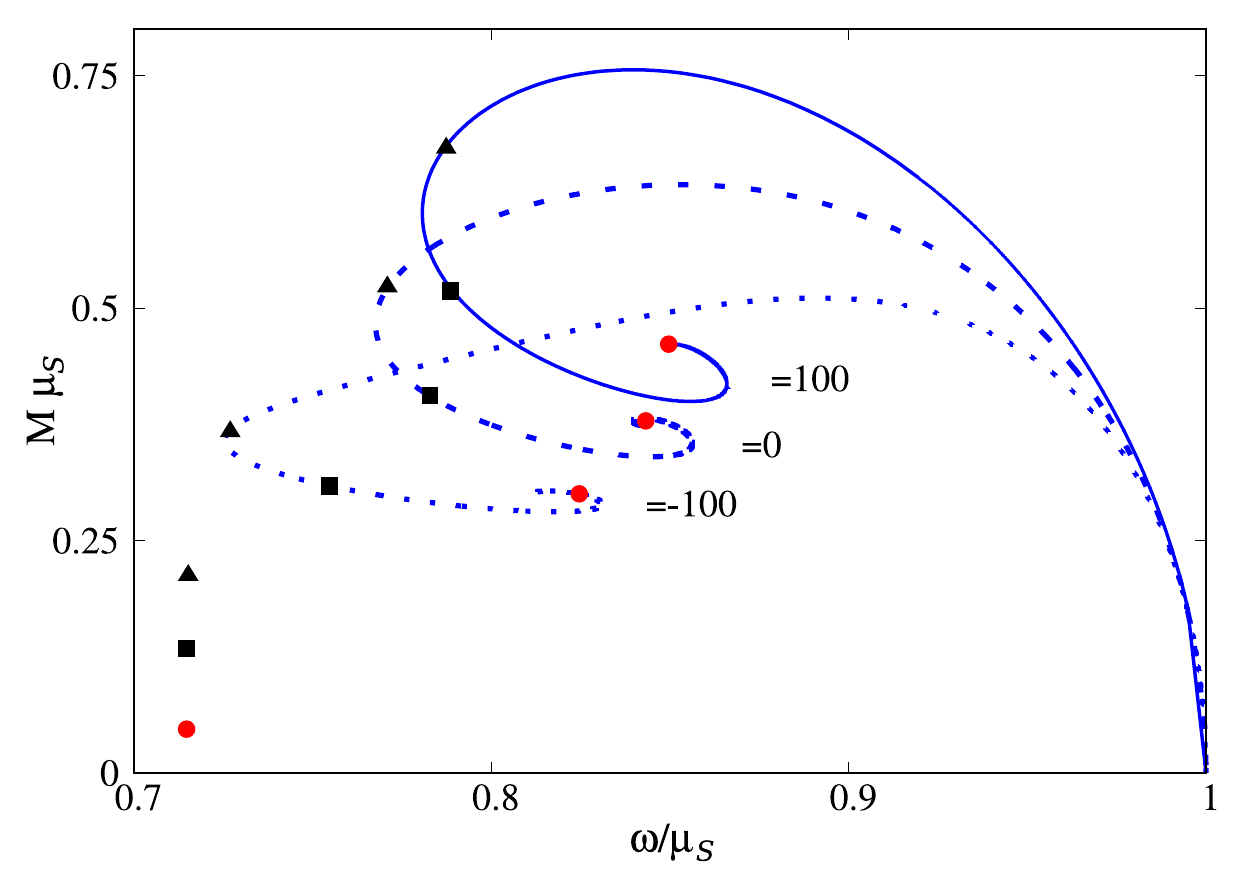}
		  \begin{picture}(0,0)
			\put(20,132){$\xi _{\rm min}$}
			\put(75,48){$\chi(\xi _{\rm trans})$}
			\put(110,0){\small{$\lambda $}}
			\end{picture}
		 \includegraphics[scale=0.65]{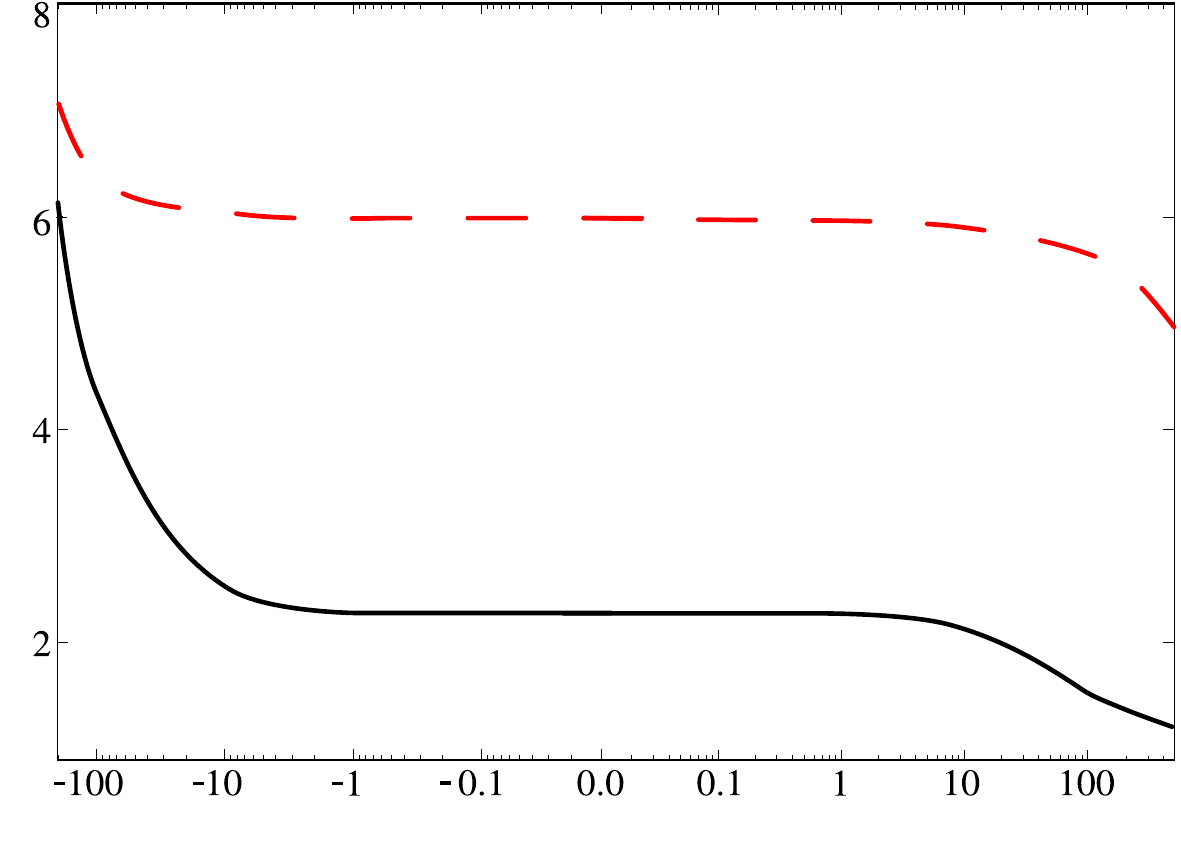}
	 	 \caption{(Left) Domain of existence of the self-interacting scalar BSs with the potential~\eqref{poly}, $\gamma=0$ and  three different values of $\lambda$: (solid) $\lambda  =100$; (dashed) $\lambda = 0$ or mini-BSs; (dotted) $\lambda =-100 $. (Right) $\xi_{\rm min}$ (dashed red line) and $\chi(\xi_{\rm trans})$ (solid black line) as a function of $\lambda$.}
	 	\label{F2}
		\end{figure}

Let us first consider the  polynomial self-interaction with the quartic term only ($\gamma=0$). The domain of existence for three values of $\lambda=\{-100, 0 , 100\}$ can be observed in Fig.~\ref{F2} (left).   The $\lambda=0$ case corresponds to the standard (scalar) mini-BSs. This model exemplifies generic behaviours observed for scalar BSs with up to quartic self-interactions  and PSs without self-interactions. In fact,  in the spherical case, scalar and vector mini-BSs have been qualitatively similar in the generality of their physical and phenomenological properties studied  in the literature so far,  only with quantitative differences.\footnote{In the rotating case,  however, a major dynamical difference was exhibited in~\cite{Sanchis-Gual:2019ljs}.} But in this paper a qualitative difference with phenomenological impact will be observed.

	The domain of existence of all studied BSs in Fig.~\ref{F2} (left)  shows a spiral behaviour starting at $\omega/\mu_S =1$  and $M \mu_S =0$,  corresponding  to the Newtonian limit wherein the stars are less compact. Starting from this limit, the BSs first increase (decrease) the ADM mass (frequency) until they reach a maximum mass ($M_{\rm max}$) at $\omega _{\rm crit}$. Then the mass decreases until the minimum frequency, which completes the, so called, first branch. After the backbending of the curve, there is a second branch. Several branches are observed, each ending on a backbending of  the curve, forming a   spiral.  The first branch, up to  $M_{\rm max}$, corresponds to the perturbatively stable BSs solutions~\cite{gleiser1988stability,gleiser1989gravitational}. This is a common feature for spherical, fundamental BSs solutions in the literature, also for PSs~\cite{brito2016proca}. We will see, however, that a qualitatively different behaviour emerges in the axionic case discussed below. In all models herein, except the latter, we refer to the \textit{perturbatively stable} BSs as the solutions in the part of the first branch that lies between  the Newtonian limit and the maximal mass solution. The  discussion in the remaining of this section applies to these ``standard" models and will be reconsidered in Section~\ref{S52} for the case of the axionic model.

Moving along the spiral, the ADM mass ($M$) and frequency ($\omega$)  undergo oscillations, as does the compactness - see~\cite{cunha2017lensing} for a plot in the case of (scalar) mini-BSs. The field amplitude at the  origin ($\varphi_0$ for the scalar case and $f_0$ for PSs), on the other hand, grows monotonically.  The latter is, therefore, a parameter that uniquely labels the solutions along the spiral. Thus, we define the ratio between the field amplitude  at  the origin ($\varphi _0,\ f_0$) of a given solution along the spiral, denoted as $x$, and the field amplitude  at the origin of  the  maximal mass  solution,
	\begin{equation}
	 \chi (x)\equiv \frac{\varphi _0 ( x ) }{\varphi _0 (M_{\rm max})} \ \ \ [{\rm{scalar}}] \qquad \textrm{or} \qquad \chi (x)\equiv \frac{f_0 ( x ) }{f_ 0 (M_{\rm max})} \ \ \ [{\rm{vector}}]\ ,
	\end{equation}
as an indicator of how  close the $x$  solution  is from  the perturbative stability crossing  point (which is the  maximal  mass solution).  In other words, 
$\chi(x)>1$ means the solution is  \textit{perturbatively unstable}.

Let us now turn our attention to the LRs and TCOs. In Fig.~\ref{F2} (left) the first ultracompact BS is denoted by a  red circle for each of the three values of $\lambda$ used. For $\lambda=0$ this solution occurs in the third  branch~\cite{cunha2017lensing}. Introducing self-interactions, this does not change and solution remains  in the perturbatively unstable region with $\chi >1$.

Consider now the TCOs.  For small values of $\varphi_0$, the areal radius of the maximum of $\Omega$, $R_\Omega$, is zero. Then, moving along the spiral, at a critical value of $\varphi_0$, there is a \textit{transition}: $R_\Omega$ starts to move away  from the origin. The transition solution, at  which $R_\Omega$  starts to move away  from the centre, is denoted by a triangle on each  of the curves in Fig.~\ref{F2} (left). We can see this solution always has $\chi>1$: it is in the perturbatively unstable region.  

As discussed in the introduction, $R_\Omega$ was seen to play the role of a BH ISCO in the simulations in~\cite{olivares2018tell}. However, the question arises, does it provide a similar scale, for a BS and a Schwarzschild BH with the same ADM mass? To  analyse this possibility, we introduce the ratio
	\begin{equation}
\xi \equiv\frac{R_{\rm ISCO}}{R_\Omega} \  .
\label{xi}
	\end{equation}

For each value of $\lambda$,  we observe that there is a minimum value of $\xi$, denoted $\xi_{\rm min}$, which occurs for a solution with  $\chi(\xi_{\rm min})>\chi(\xi_{\rm trans})>1$,  where $\xi_{\rm trans}$ represents the transition solution wherein $R_\Omega$  starts to depart  from the origin. The solution with $\xi_{\rm min}$  is denoted by a black square on each  of the curves in Fig.~\ref{F2} (left).

To summarise, 	Fig.~\ref{F2} (left) shows that, in this model, BS solutions with $R_\Omega\neq 0$ only occur after $M_{\rm max}$ and thus are perturbatively unstable. The transition solution responds to a positive (negative) coupling approaching (moving  away) from the stable  branch. We can observe from  Fig.~\ref{F2} (right) that $\chi(\xi_{\rm  trans})$ approaches unity for the largest positive values  of $\lambda$, but does not quite reach it within the values of $\lambda$ explored. In this limit, the minimum value of $\xi_{\rm min}$ is still larger than 5.

	From the data presented, quartic self-interactions make BSs  have their transition solution closer to the stable region (than mini-BSs) but not quite reaching it. The  first ultralight compact scalar BSs, on the other hand, does not  approach noticeably the stable region.

	\subsection{The polynomial self-interaction:  $\gamma\neq 0$}\label{S511}
%
	
\begin{figure}[h]
		 \centering
		 	\begin{picture}(0,0)
		    \put(41.5,21){$1^{st}\ LR$}
			 \put(41.5,39){$\xi _{\rm min}$}
			 \put(41.5,57){$\xi _{\rm trans}$}
			 \put(88,94.5){\small{$\gamma $}}
			 \put(108,85){\small{$\gamma $}}
			 \put(118,74){\small{$\gamma $}}
			\end{picture}
		 \includegraphics[scale=0.63]{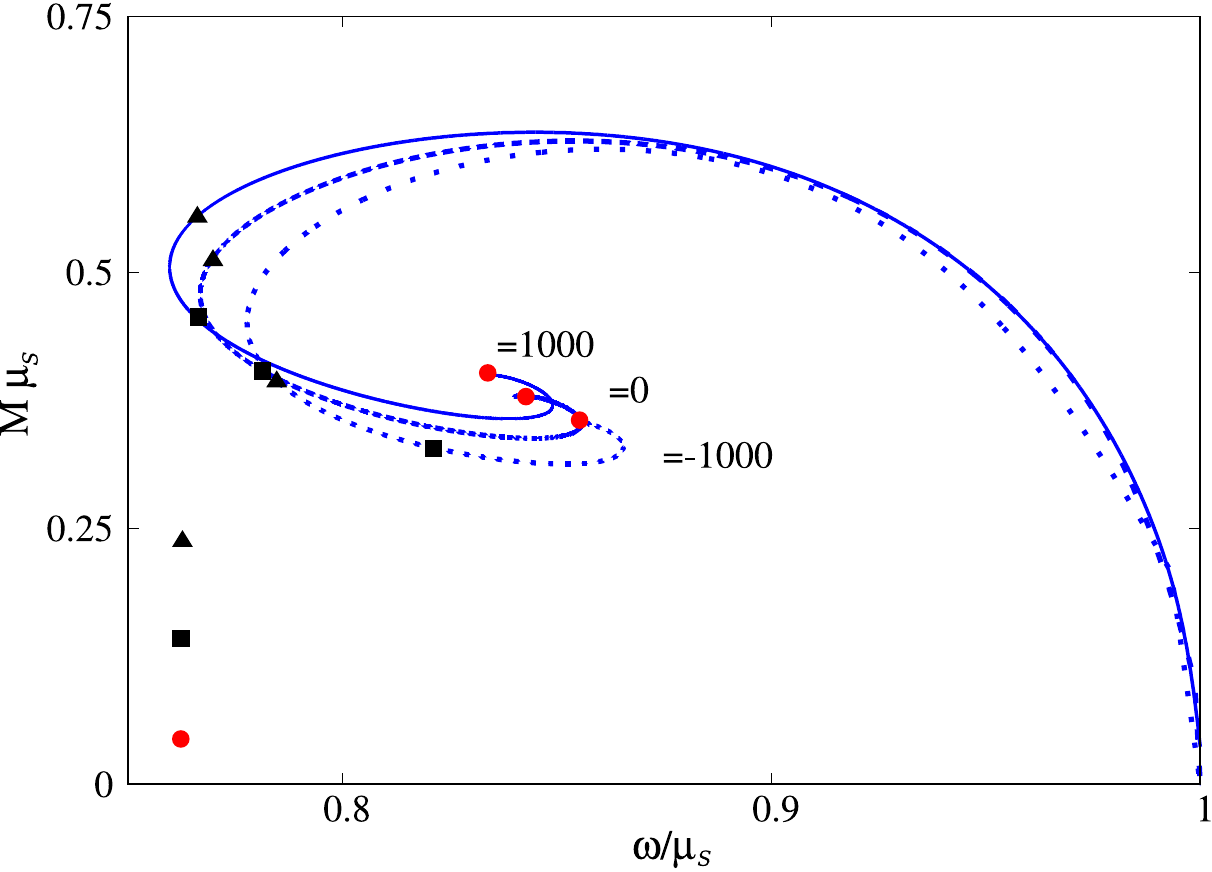}
		  \begin{picture}(0,0)
			\put(30,138){$\xi _{\rm min}$}
			\put(90,60){$\chi(\xi _{\rm trans})$}
			\put(116,0){\small{$\gamma $}}
			\end{picture}
		 \includegraphics[scale=0.63]{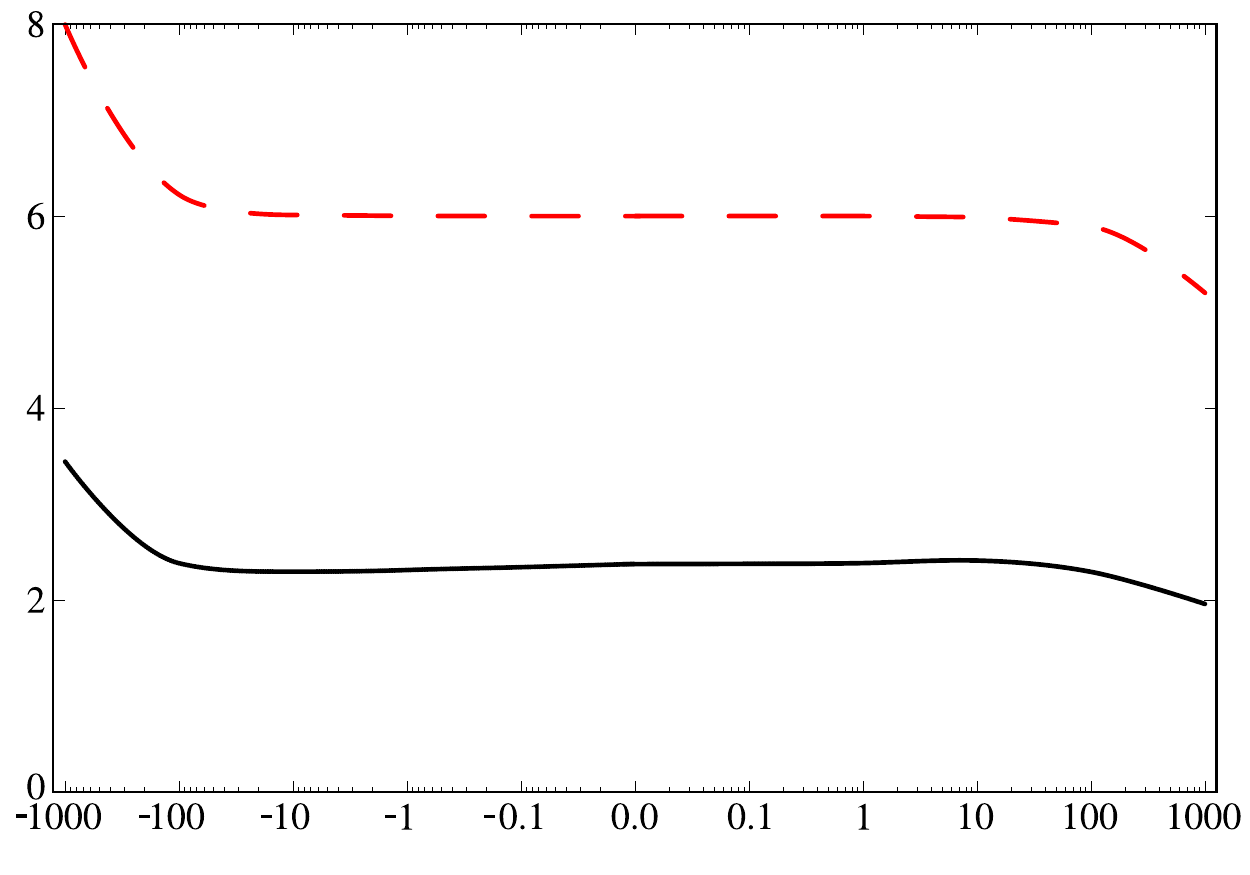}
	 	 \caption{(Left) Domain of existence of the self-interacting scalar BSs with the potential~\eqref{poly}, $\lambda =1.0$ and  three different values of $\gamma$: (solid) $\gamma  =1000$; (dashed) $\gamma = 0$; (dotted) $\gamma =-1000 $. (Right) $\xi_{\rm min}$ (dashed red line) and $\chi(\xi_{\rm trans})$ (solid black line) as a function of $\gamma$.}
	 	\label{F4}
		\end{figure}
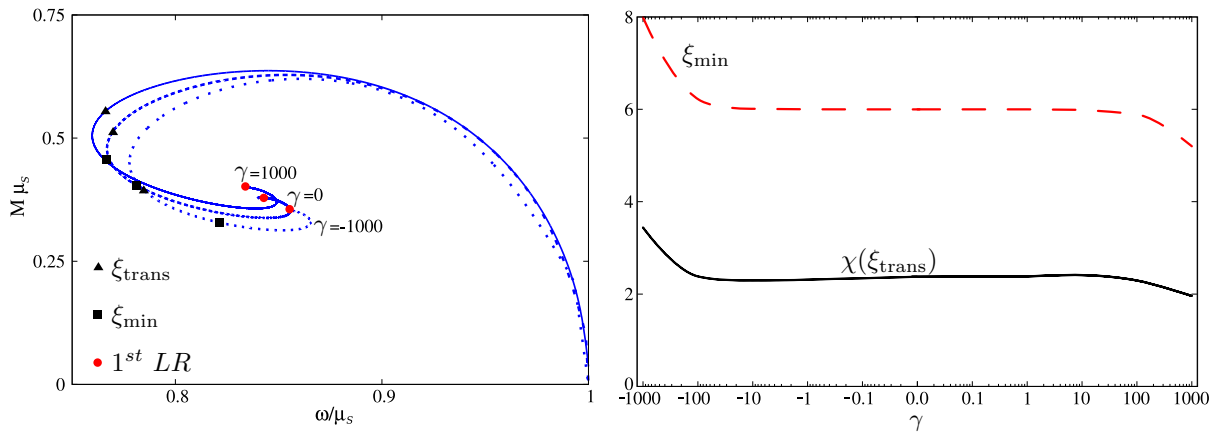

	We now consider the  polynomial self-interaction with both the quartic and sextic terms ($\gamma\neq 0\neq \lambda$). From the data presented before and in an attempt to monitor the behaviour of the sextic coupling, we will consider $\lambda=1.0$ and change the value of $\gamma$. Again, for each value of $\gamma$, we will have a family of solutions with an associated domain of existence, as represented in Fig.~\ref{F4} (left).

The trends observed in Fig.~\ref{F4} are qualitatively similar to the ones seen in the previous  section, for $\gamma=0$ and varying $\lambda$.
In this case, however, both the first ultracompact solution and the transition solution  respond to a positive (negative) coupling approaching (moving  away) from the stable  branch. But again, these solutions  do not reach  the  perturbative stability region for the values of $\gamma$ explored.

This analysis and the one  in the previous subsection, suggest that a simultaneous increase in both $\lambda$ and $\gamma$ may bring $\xi _{\rm min}$ and $\chi (\xi _{\rm trans})$  closer to unity. To  test this hypothesis, we have computed scalar BSs  with $\lambda =100$ and $\gamma =1000$, corresponding to the two largest values that our code supported with trustable results. We obtained that   the transition  occurs closer to the maximum mass, $\chi (\xi _{\rm trans})=1.51$, but the $R_\Omega$ is still fairly below the ISCO radius of the comparable BH: $\xi _{\rm min}=5.09$. Concerning the first ultracompact solution, we obtained $\chi (LR)=6.58$, still far from the stability region. 
	
	From the data presented, the inclusion of a $6^{th}$ order self-interaction term does not lead to either ultracompact scalar BSs or such stars with $R_\Omega\neq 0$ in the perturbatively stable region  although one observes a trend that could approach the  stability region. We also observe that a $6^{th}$ self-interaction term has a smaller impact in the overall scalar BS solutions than the quartic coupling, which is mainly associated to the fact that the scalar field is never larger than unity.
%
\subsection{The axionic self-interaction}\label{S52}
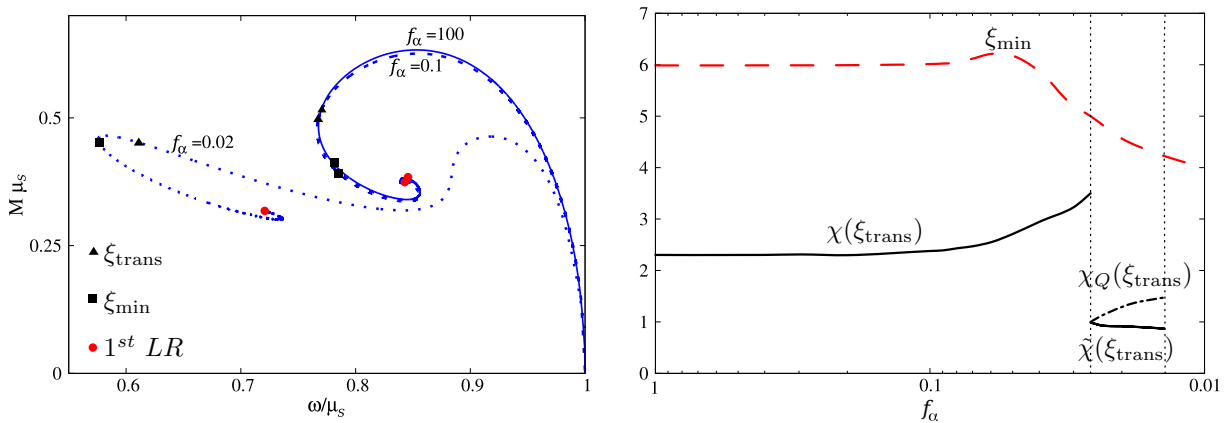
\begin{figure}[h]
			 \centering
			\begin{picture}(0,0)
		    \put(41,25){$1^{st}\ LR$}
			 \put(41,43){$\xi _{\rm min}$}
			 \put(41,61){$\xi _{\rm trans}$}
			\end{picture}
			\includegraphics[scale=0.63]{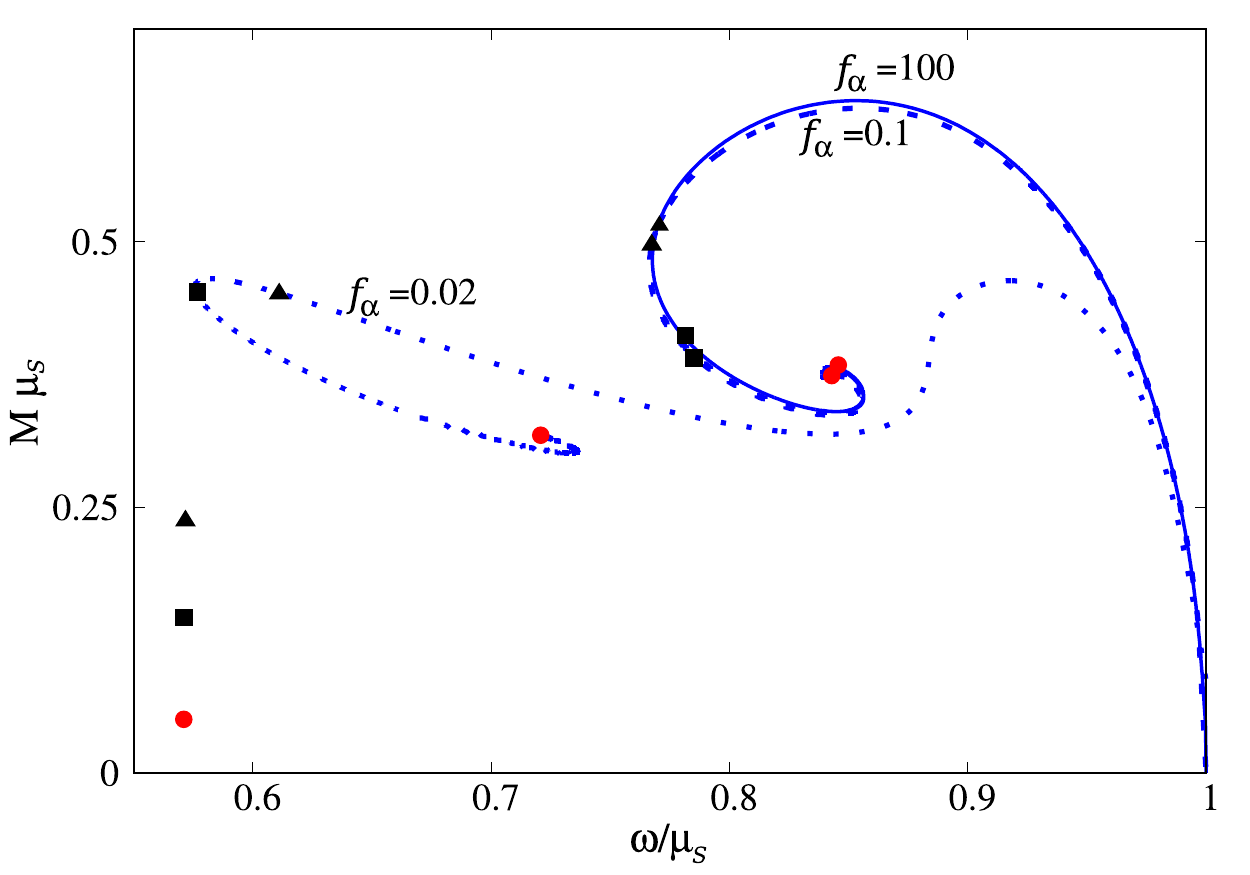}
		  \begin{picture}(0,0)
			\put(140,143){$\xi _{\rm min}$}
			\put(80,70){$\chi(\xi _{\rm trans})$}
			\put(175,25){$\tilde{\chi} (\xi _{\rm trans})$}
			\put(175,53){$\chi _{Q} (\xi _{\rm trans})$}
			\end{picture}
			\includegraphics[scale=0.63]{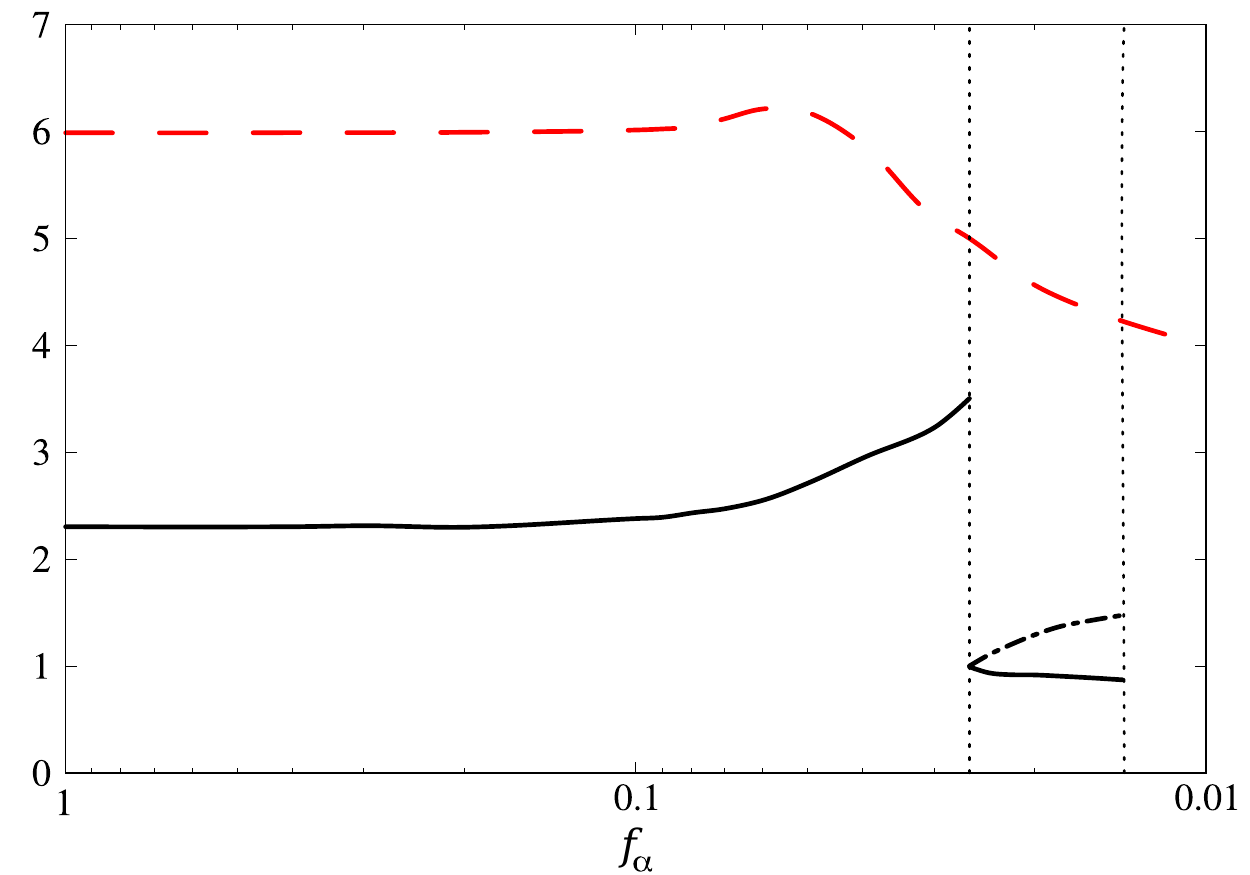}
	 		 \caption{(Left) Domain of existence of the self-interacting scalar BSs with the axionic potential~\eqref{axionp} and three different values of $f_\alpha$: (solid) $f_\alpha =100$; (dashed) $f_\alpha =0.1$; (dotted) $f_\alpha =0.02 $. 
(Right) $\xi_{\rm min}$ (dashed red line), $\chi(\xi_{\rm trans})$  (solid black line) and $\chi _Q (\xi_{\rm trans})$ (dotted-dashed black line) as a function of $f_\alpha$. {The left/right black vertical dotted lines define the value  of $f_\alpha$ for which  the second/third maximum first appears. The region to the right of the first vertical line encompasses BS solutions  with a relativistic stable branch.}}
	 		 \label{F5}
	\end{figure}

%
As our last example of  spherical scalar BSs we use the axionic potential, first considered in~\cite{guerra2019axion}. The domain of existence for three different values of the coupling constant $f_\alpha=\{ 100,0.1,0.02\}$ is represented in Fig.~\ref{F5} (left). 

For a large value of $f_\alpha$ one recovers the scalar mini-BSs. Then, as $f_\alpha$ decreases, qualitatively different properties start to emerge. Most notably, a second local maximum  for the mass appears {(see Fig.~\ref{F5} (right) left vertical line)} that, for some value of $f_\alpha$ (close to $f_\alpha =0.02$), becomes the global maximum. This qualitatively different domain of existence impacts on the stability region as we now discuss. But before that, let us introduce two new quantities:	
	\begin{equation}
	 \tilde{\chi} (\xi_{\rm trans}) =\frac{\varphi _0 (\xi_{\rm trans})}{\varphi _0 (2^{\rm nd} M_{\rm max})}\ , \qquad  \qquad \chi _Q(\xi_{\rm trans})=\frac{\varphi _0 (\xi _{{\rm trans}})}{\varphi _0 (Q=M)}\ ,
	\end{equation}
	where $\tilde{\chi}	(\xi_{\rm trans})$ $\big[\chi _Q(\xi_{\rm trans}) \big]$ gives a measure of the distance between the solutions where $R_\Omega$ depart from the origin and the second maximum [solution for which the Noether charge equals the ADM mass $Q\mu_S=M$].
	
\begin{figure}[h]
			 \centering
			\includegraphics[scale=0.66]{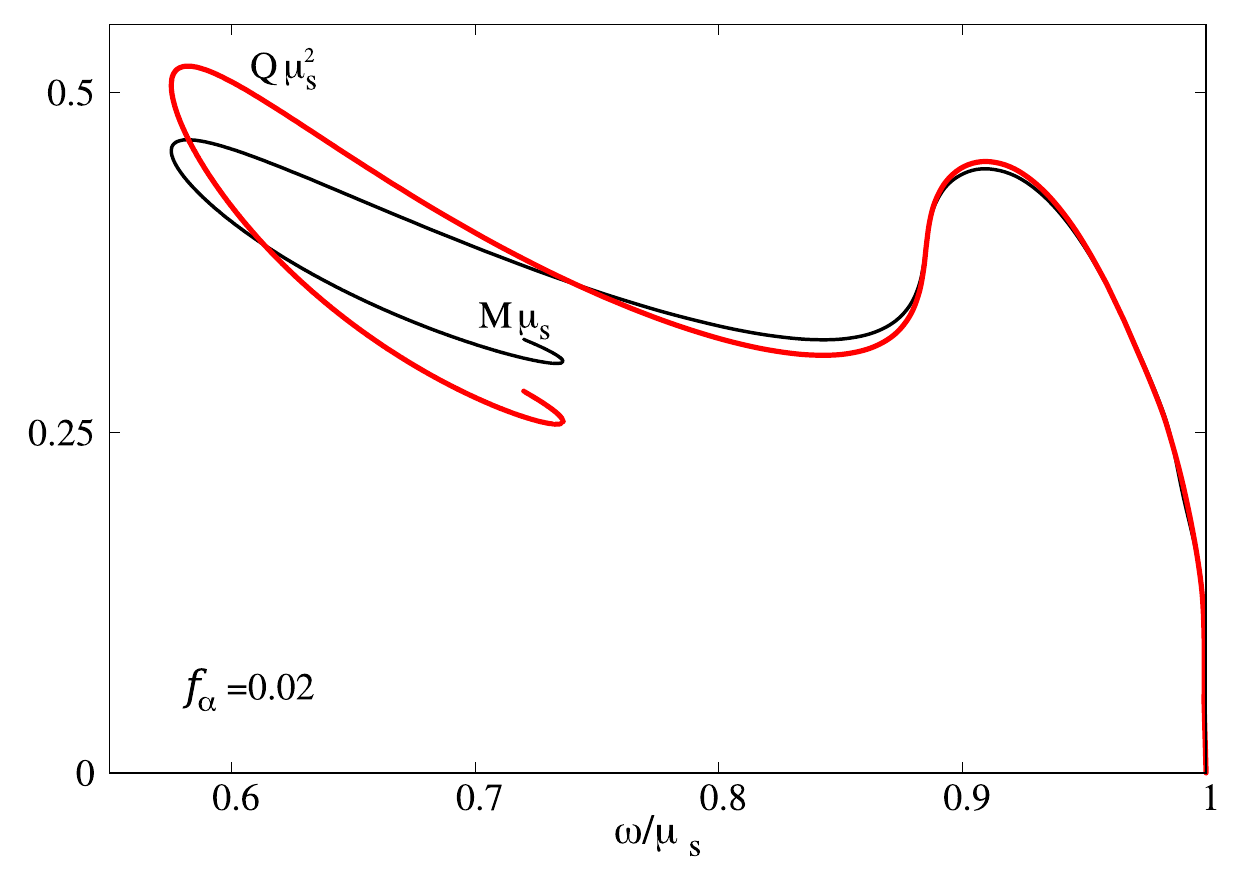}
    \includegraphics[scale=0.30]{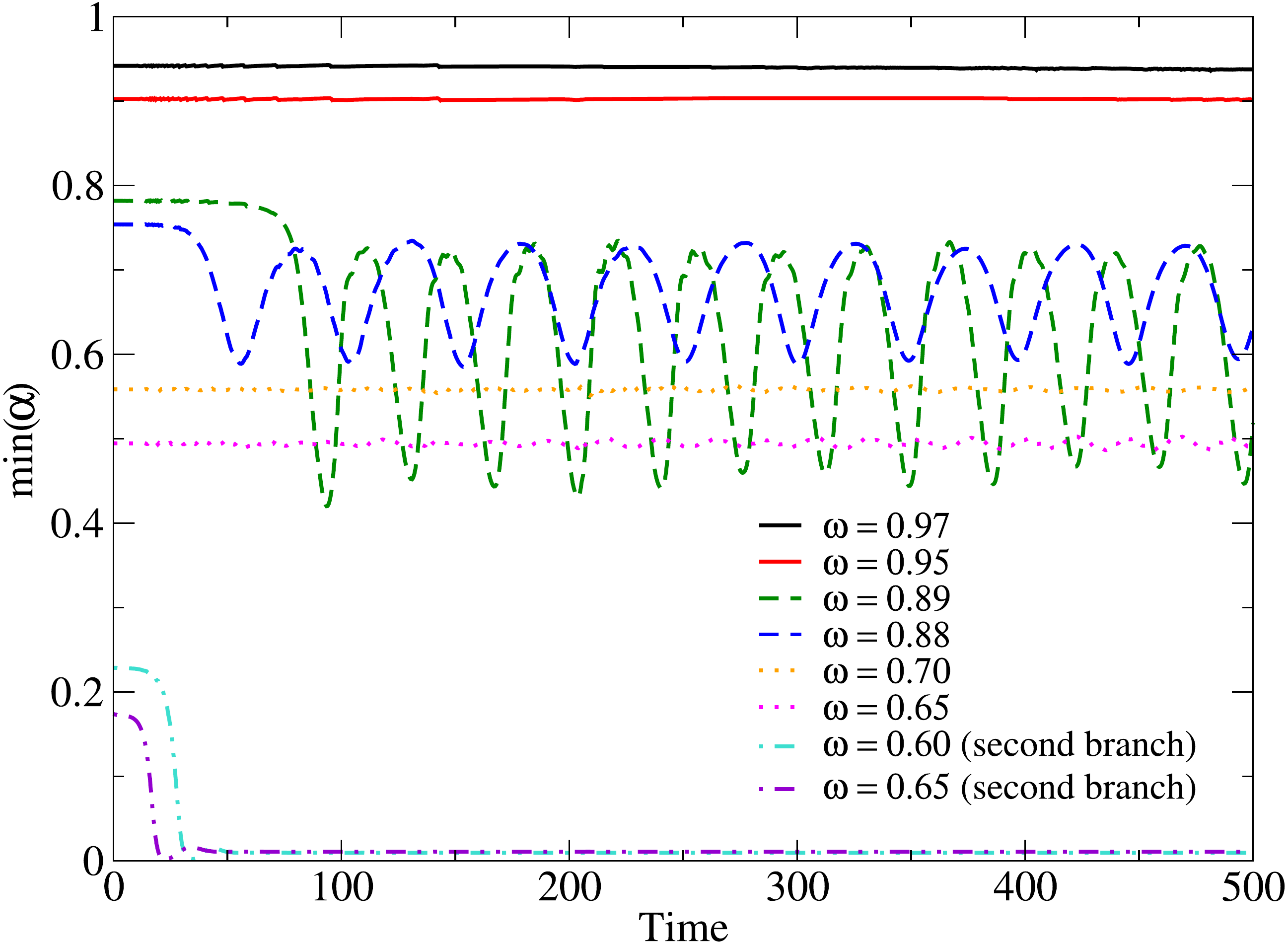}
			\includegraphics[scale=0.31]{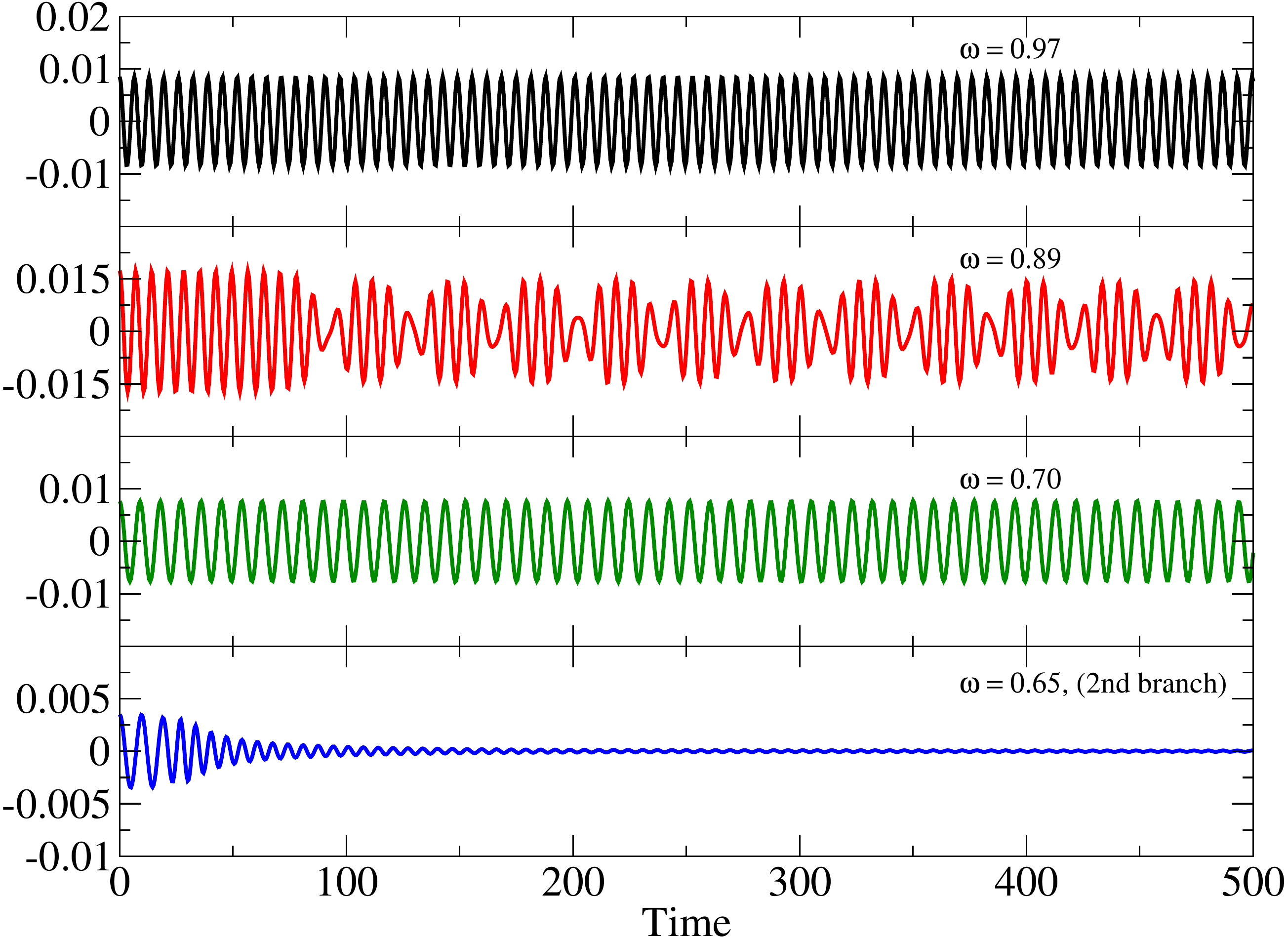}
                         \includegraphics[scale=0.31]{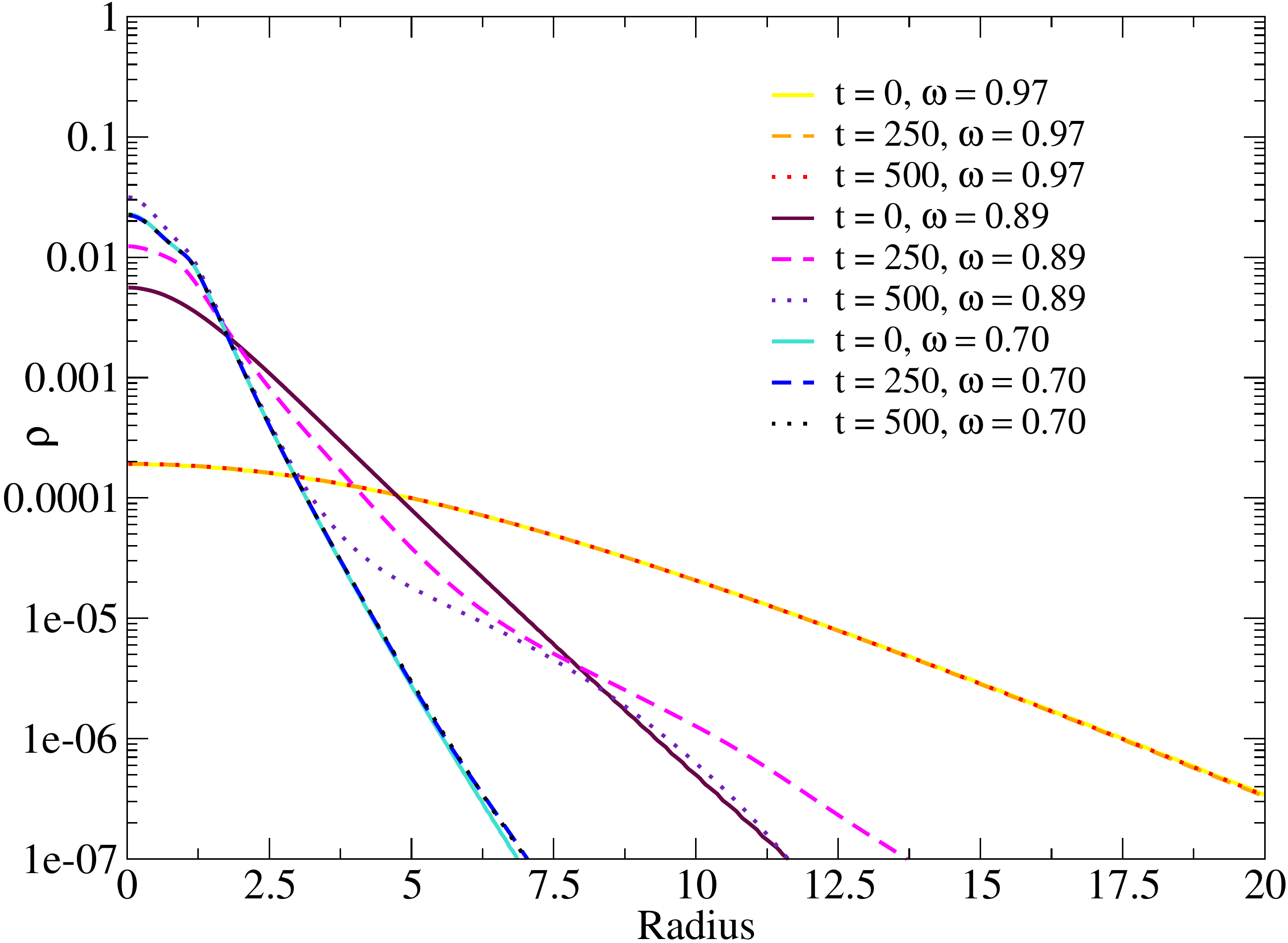}
                     
	 		 \caption{(Top left) Domain of existence of the self-interacting scalar BSs with the axionic potential~\eqref{axionp} for both the ADM mass and Noether charge.  Evolution of the minimum of the lapse (top right) and scalar field at an illustrative observation radius (bottom left) for several models. (Bottom right) Evolution of the radial profile for three illustrative models.}
	 		 \label{F6}
	\end{figure}

In all previous cases of BSs it has been argued that solutions in the $M-\omega$ domain of existence  after the global maximum are unstable. The existence of such unstable modes originating  at such critical point of the ADM mass has  been explicitly shown by perturbation theory studies in different works, $e.g.$~\cite{kusmartsev1991stability,gleiser1988stability,gleiser1989gravitational,brito2016proca}. However, could there be other stability regions further into the spiral? In~\cite{Kleihaus:2011sx} it was argued that catastrophe theory arguments suggest that BS models  with a  domain of existence akin to that for $ f_ \alpha=0.02$ in Fig.~\ref{F5}, have two stable branches: the \textit{Newtonian} one between the maximal frequency and the first (local) maximum of the mass, and the \textit{relativistic} one between   the first local minimum of the mass and the second local (which can be the global) maximum. {The second region occurs to the  right   of  the leftmost vertical dotted line in Fig.~\ref{F5} (right)}. To test this conclusion we have performed fully non-linear numerical evolutions, using the same setup and code as described in~\cite{sanchis2016explosion,sanchis2016dynamical,escorihuela2017quasistationary,cunha2017lensing,herdeiro2018spontaneous}, of different BSs corresponding to our axionic model~\eqref{axionp}. The code uses spherical coordinates under the assumption of spherical symmetry employing the second-order Partially Implicit Runge-Kutta (PIRK) method developed by \cite{montero2012bssn,cordero2012partially,cordero2014partially}. Our results are exhibited in Fig.~\ref{F6}.

The results in Fig.~\ref{F6} support that there are indeed two disjoint stable branches of solutions. In the top left panel we show both the ADM mass and the Noether  charge $Q$ for  $f_\alpha\simeq 0.02$. The regions where $Q\mu_S<M$ $(\chi _Q <1)$ correspond to solutions with energy excess, which are not energetically stable. But the first (top) branch of solutions between the minimum of the  frequency and the crossing point between $M$ and $Q\mu_S$, corresponding to $\omega/\mu_S\in [0.582, 0.753]$ are stable.  This is the relativistic stable branch. On the other  hand, the Newtonian stable branch corresponds  to  $\omega/\mu_S\in  [0.92, 1]$.\footnote{In all the Newtonian branch, $Q\mu_S/M>1$ with the crossing point at $\omega/\mu_S\simeq 0.885$.}  The stability in these branches is corroborated by the analysis in the top right panel, exhibiting the minimum of the lapse function $\alpha$ as a function of time, during the evolutions.  Solutions in these stable branches have an approximately constant lapse, as illustrated by $\omega/\mu_S=0.97,0.95$ (Newtonian branch) and  $\omega/\mu_S=0.70,0.65$ (relativistic branch),\footnote{The small oscillations seen  come from the interpolation, the different resolutions of the two grids in the two refinment levels used in the simulations and the outer boundary.} whereas the  solutions in the frequency range in between exhibit large oscillations, as illustrated by $\omega/\mu_S=0.88, 0.89$. These solutions, however, do not decay into BHs. Indeed, two solutions in the second (bottom) branch of the left top panel,   with  $\omega/\mu_S=0.60, 0.65$, that collapse to BHs are  also shown, for comparison. The bottom left panel  shows the scalar field extracted at an illustrative radius, and corroborates the different qualitative behaviour between the stable and unstable branches. The bottom right panel exhibits the energy density of the BSs, as a function of the radius,  for the different models and for different times during the evolution. One can observe that: for  $\omega/\mu_S=0.97$ (Newtonian stable branch) and for $\omega/\mu_S=0.7$ (relativistic  stable branch) the radial profile does  not change in time; but for $\omega/\mu_S=0.89$ the profile changes and the solution approaches the profile of the (more compact) solution in the relativistic stable branch. We conclude that the unstable models in between the two stable branches do not collapse (even if perturbed) but rather migrate to the relativistic stable branch, where the BSs are more compact.

	{If we see now Fig.~\ref{F5} (right) we observe a region between $f_{\alpha} = [0.026, 0.016]$ where solutions can be both relativistic stable $ \tilde{\chi} (\xi_{\rm trans}) <1$ and energetically stable $\chi _Q (\xi_{\rm trans})>1$. Below $f_\alpha = 0.016$ the domain of existence becomes more complex -- with further local maximums -- and, while it could exhibit further stable regions, the numerical results are not precise enough to further explore that region.}

With the demonstrated evidence for a new stable branch we can extract our main conclusion from Fig.~\ref{F5} (left): for $f_\alpha=0.02$ the transition point is already in the relativistic stable branch. Thus, unlike the previously analysed models, there are dynamically robust axionic BSs with  $R_\Omega\neq 0$. However, we can see from Fig.~\ref{F5} (right) that the $\xi_{\rm min}$ is still not one, but further decreasing $f_\alpha$ it exhibits a decreasing trend.

\section{Proca Stars}\label{S5}
%
	Let us now consider the self-interacting PS model, with the potential~\eqref{vp}, first considered in~\cite{minamitsuji2018vector}. Setting the self-interactions coupling $\lambda_P$ to zero, this model yields mini-PSs  as  solutions~\cite{herdeiro2017asymptotically,sanchis2017numerical,brito2016proca}.

\begin{figure}[h!]
		 \centering
		 	\begin{picture}(0,0)
			 \put(41,25){$1^{st}\ LR$}
			 \put(41,41){$\xi_{\rm min}$}
			 \put(41,56){$\xi^{2^{nd}} _{\rm trans}$}
			 \put(102,25){$\xi ^{1^{st}} _{\rm trans}$}
			 \put(102,41){$\xi =1$}
			 \put(155,66){\small{$\lambda _P =0.01$}}
			 \put(135,80){\small{$\lambda _P =0$}}
			 \put(120,98){\small{$\lambda _P =-1.0$}}
			\end{picture}
		 \includegraphics[scale=0.63]{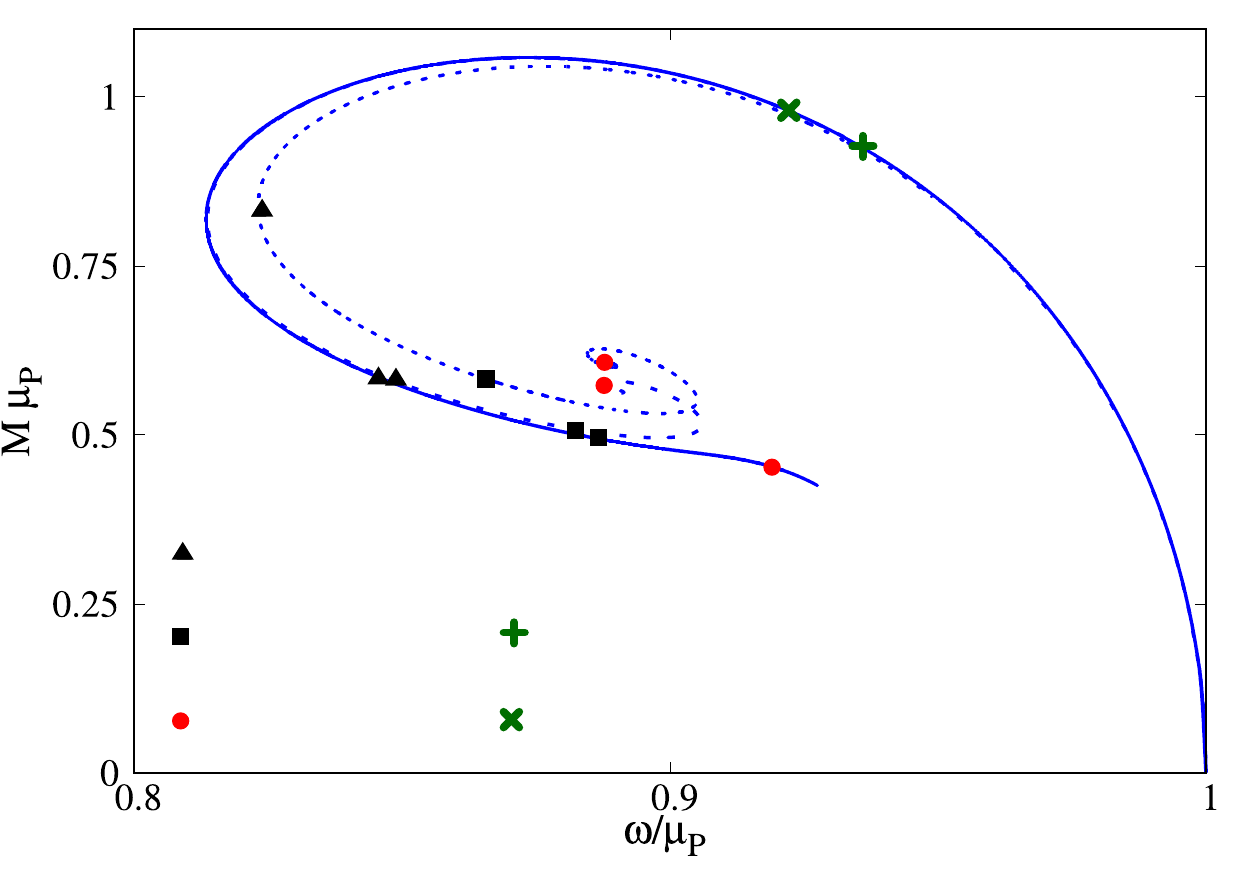}\hfill
		  \includegraphics[scale=0.63]{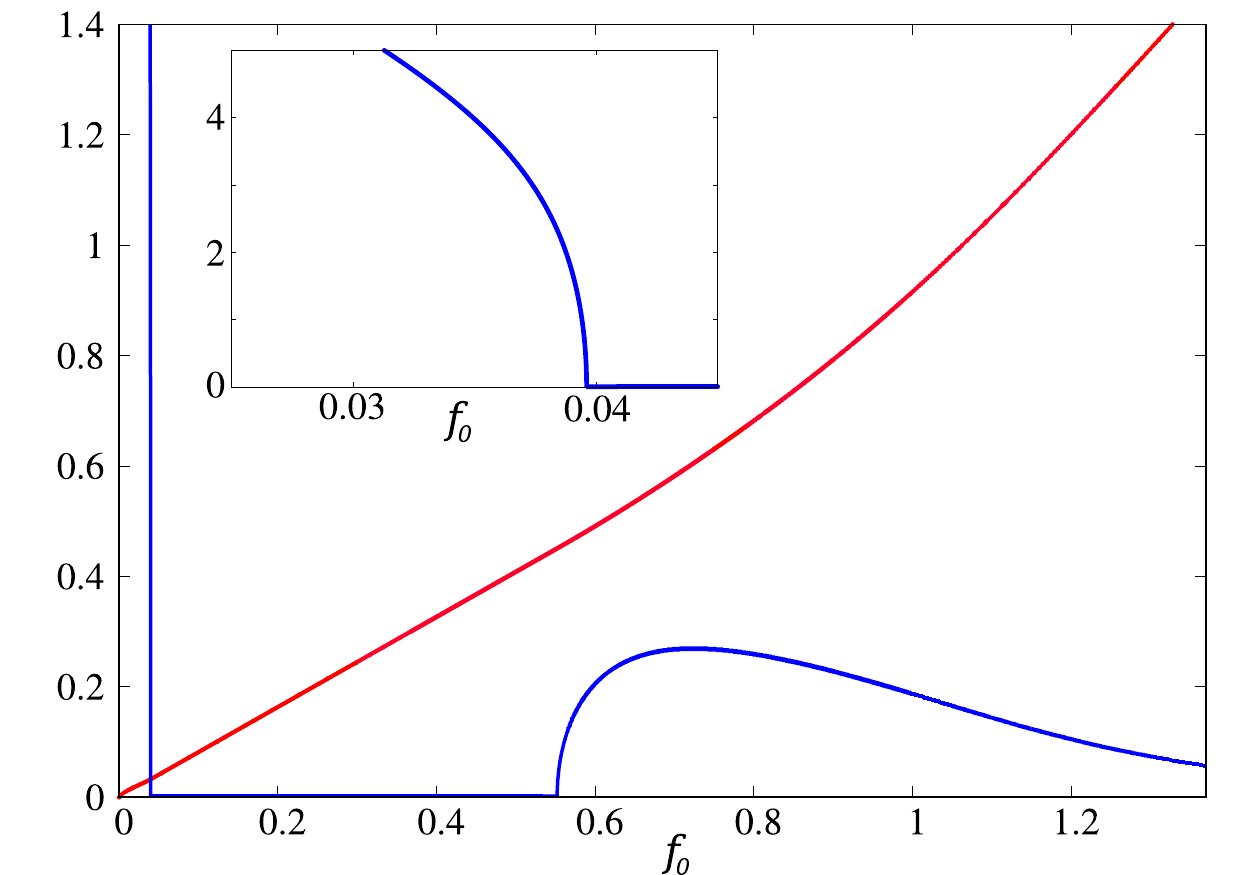}
		 	\begin{picture}(0,0)
		 	 \put(-50,70){$\lambda _P = 0$}
			 \put(-150,120){$R_\Omega$}
			 \put(-50,30){$R_\Omega$}
			 \put(-70,130){$\Omega _{\rm max}$}
			\end{picture}
	 	 \caption{(Left) Domain of existence of the self-interacting PSs with  three different values of the self-interaction coupling: (solid) $\lambda _P =0.01$; (dashed) $\lambda _P = 0$ or mini PSs; (dotted) $\lambda _P =-1.0 $. (Right) Areal radius of the maximal angular velocity along TCOs, $R_\Omega$ (blue solid line)  and the corresponding value of the angular velocity $\Omega _{\rm max}$ (red solid line), as a function of the Proca field amplitude at the origin $f_0$ for a mini-PS.}
	 	\label{F01}
		\end{figure}
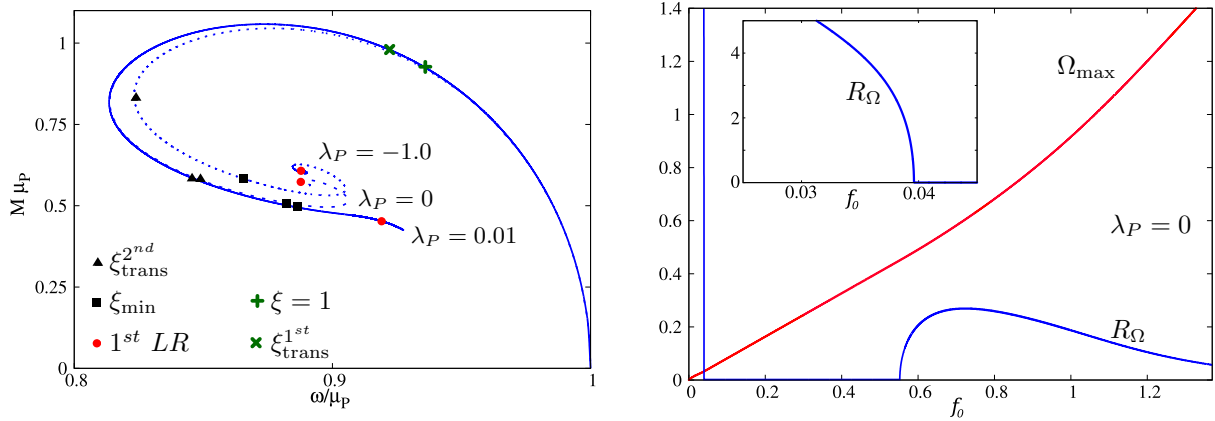

	The domain of existence of all studied PSs  shows a spiral behaviour starting at $\omega/\mu_P =1$  and $M \mu_P =0$,  corresponding  to the less compact stars - see Fig.~\ref{F01} (left). From this limit, the PSs first increase (decrease) the ADM mass (frequency) until they reach the maximum mass, $M_{\rm max}$, at $\omega _{\rm crit}$. Then the mass decreases until the minimum frequency, which completes the first branch. After the backbending of the curve, there is a second branch. The behaviour then depends on the coupling $\lambda_P$. For $\lambda_P\leqslant 0$, several branches are observed, each ending on a backbending of  the curve, forming a   spiral.  For $\lambda _P >0$ the domain of existence is qualitatively different.\footnote{This may be understood~\cite{minamitsuji2018vector} from  the fact that for the fundamental PSs, $f(r)$ must have a node; for $\lambda _P>0$ a maximum of  $f_0$ exists compatible with this requirement
	\begin{equation}
	0<f_0\leqslant \frac{\mu _P \sigma _0}{\sqrt{\lambda _P}}\ ,
	\end{equation}
	which was numerically confirmed. For $\lambda _P <0$ there is also a critical value of $f_0$ beyond which solutions cease to exist; however it  appears to be less sensitive to the  coupling.} 
The first branch, up to  $M_{\rm max}$, corresponds to the perturbatively stable PSs solutions~\cite{brito2016proca}.

Let us now turn our attention to the LRs and TCOs. In Fig.~\ref{F01} (left) the first ultracompact PS is denoted by a  red circle for each of the three values of $\lambda_P$ used. For $\lambda_P=0$ this solution occurs at the beginning  of the fourth  branch~\cite{cunha2017lensing}. Introducing self-interactions, this solution remains  in the perturbatively unstable region with $\chi >1$.

Consider now the TCOs. Starting at $\omega/\mu_P =1$ and $M \mu_P =0$ ($f_0\rightarrow 0$), the areal radius of the maximum of $\Omega$, $R_\Omega$, is very large - see the top inset in Fig.~\ref{F01} (right). Then, moving along the spiral, at a critical value of $f_0$, there is a first \textit{transition}: $R_\Omega\rightarrow 0$ (denoted in Fig.~\ref{F01} (left) as a green cross). This occurs  for
 \[
\omega / \mu _P \approx 0.923 \ , \qquad  M \mu _P \approx 0.979 \ , \qquad \chi (\xi ^{1^{st}} _{\rm trans} )\approx 0.355 
\]
well within the stable branch, and it is rather insensitive to the coupling $\lambda_P$  in  the models  explored.

Along the sequence of solutions between the Newtonian limit and the first transition, the ratio $\xi$~\eqref{xi} varies from a large value to zero. This means that a certain PS configuration in the \textit{perturbatively stable Newtonian branch} has $R_\Omega=6M$. This solution with $\xi =1$ (denoted in Fig.~\ref{F01} (left) as a green plus) has
\begin{equation}
\label{specialone}
 \omega / \mu _P \approx 0.936 \ , \qquad M \mu _P \approx 0.925\ , \qquad  \chi (\xi=1)\approx 0.248  \ .
\end{equation}

	After the first transition, continuing along the spiral, $R_\Omega=0$ until a second transition occurs - see the bottom inset in Fig.~\ref{F01} (right). The second transition solution, at  which $R_\Omega$ moves away from the origin, is denoted by a triangle on each of the curves in Fig.~\ref{F01} (left). We can see this solution always has $\chi>1$: it is in the perturbatively unstable region and it  depends on  $\lambda_P$.  The red curve in Fig.~\ref{F01} (right) shows that the maximal value of the angular velocity, $\Omega_{\rm max}$, increases monotonically with $f_0$.

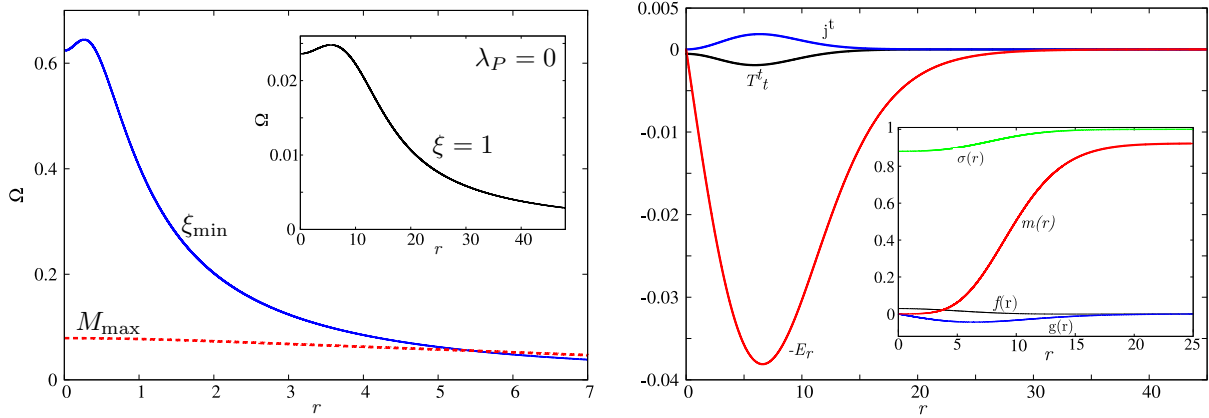
\begin{figure}[h!]
		 \centering
		  \includegraphics[scale=0.63]{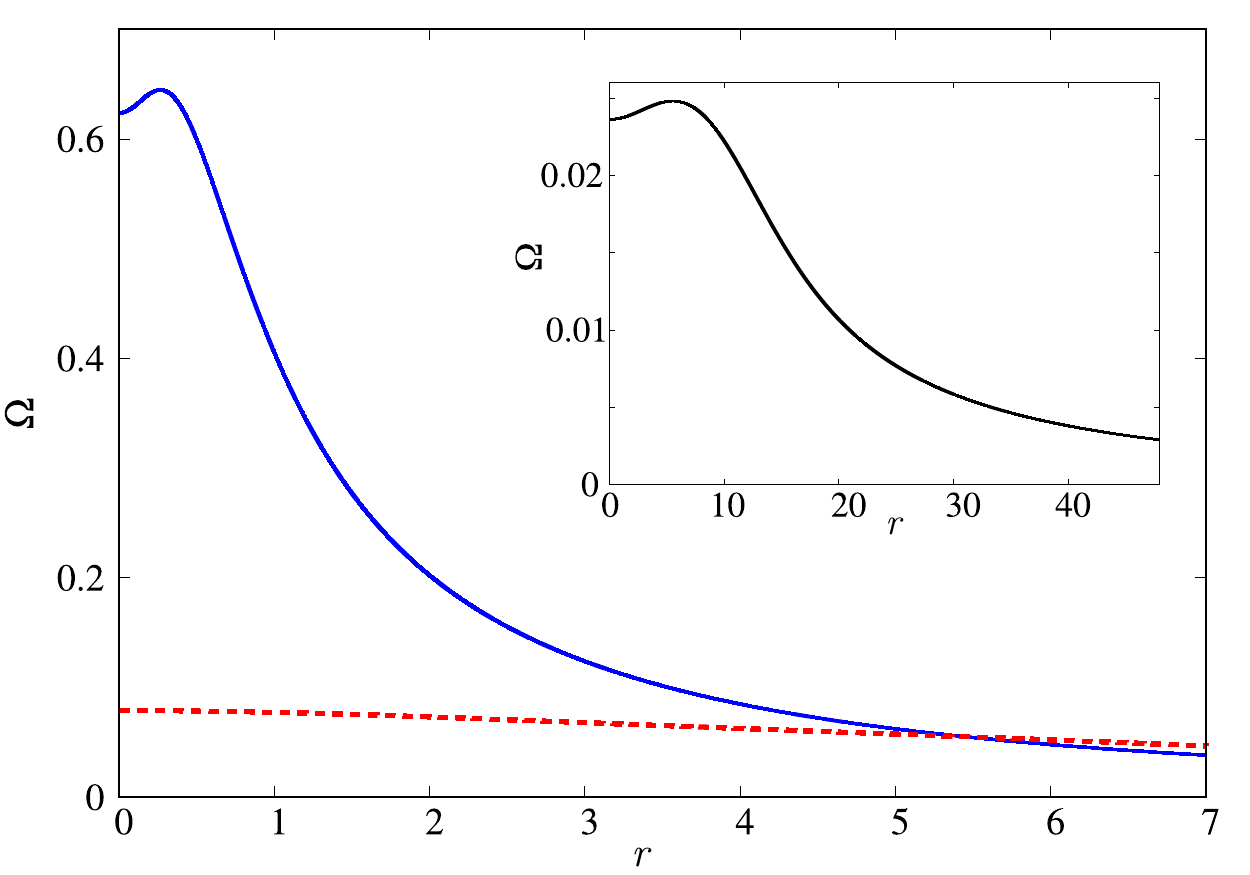}
		     \begin{picture}(0,0)
		     \put(-54,133){$\lambda _P = 0$}
			 \put(-165,70){$\xi_{\rm min}$}
			 \put(-205,33){$M_{\rm max}$}
			 \put(-70,100){$\xi =1$}
			\end{picture}
		\includegraphics[scale=0.63]{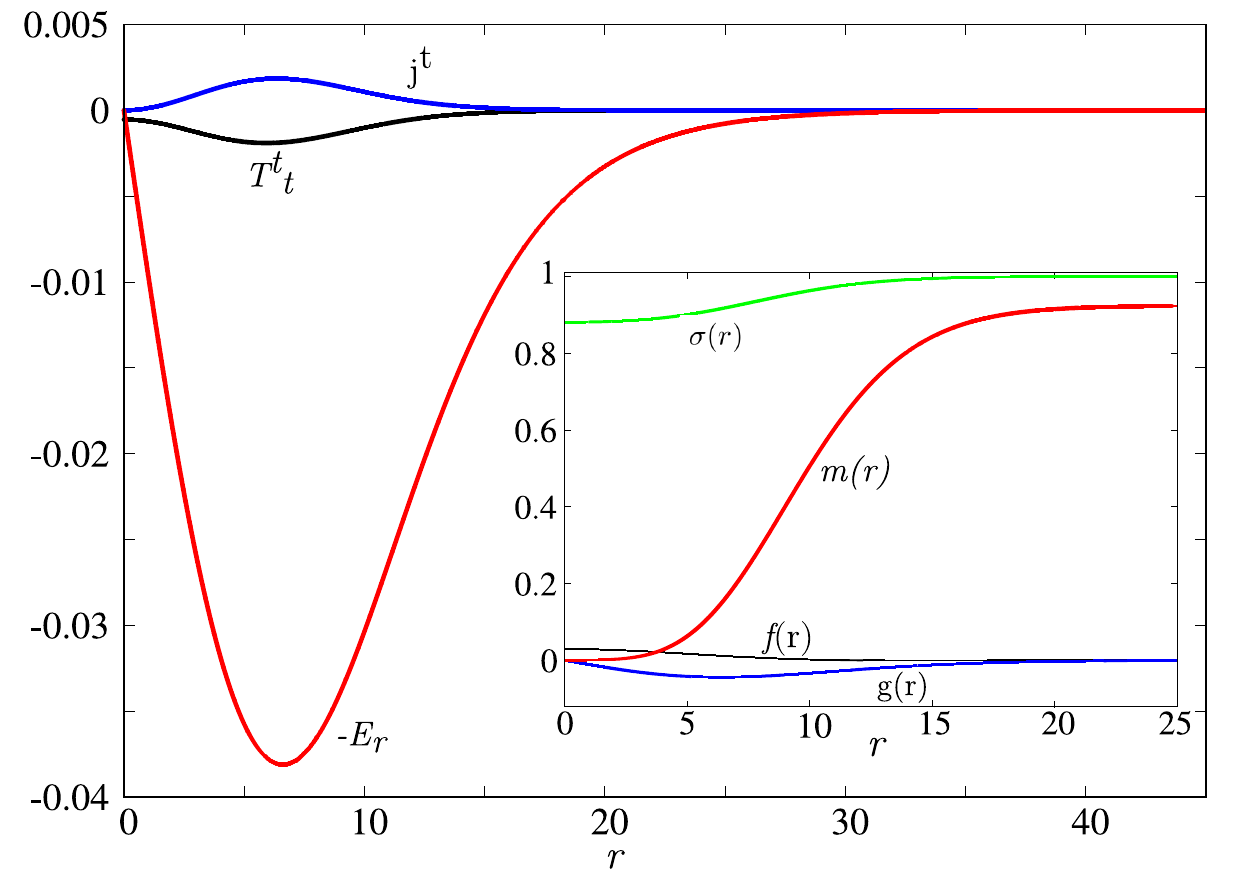}
	 	 \caption{(Left) $\Omega(r)$ for the maximal mass mini-PS (red dashed line), for the solution with minimal $\xi$ (solid blue line) and (inset) for the $\xi=1$ solution (solid black line), for $\lambda_P=0$. (Right) Radial profiles of several quantities for the ``special" solution  with $\xi=1$~\eqref{specialone}:  (main panel) energy density $T^t_t$ (see~\cite{brito2016proca} for the explicit expression), Noether charge density $j^t$~\eqref{W26} and ``electric field" $E_r=| F_{rt}|$; (inset)  metric functions~\eqref{E24} and Proca potential functions~\eqref{E25}.}
	 	\label{F0}
		\end{figure}

 In Fig.~\ref{F0} (left), 
we exhibit  the radial profile of the angular velocity for two  illustrative solutions: the red dashed line corresponds to the maximal mass solution, that has $\chi=1$, for which  $R_{\Omega}$ is  still at the origin; the blue solid  line is for a PS for which $R_{\Omega}$ is already away from the origin (after the  second transition).

As in the scalar case, after the second transition, for each value of $\lambda_P$,  we observe that there is a minimum value of $\xi$, denoted $\xi_{\rm min}$, which occurs for a solution with $\chi(\xi_{\rm min})>\chi(\xi ^{2^{nd}} _{\rm trans})>1$,  where $\xi ^{2^{nd}} _{\rm trans}$  represents the second transition solution. The solution with $\xi_{\rm min}$  is denoted by a black square on each  of the curves in Fig.~\ref{F01} (left).  

One may wonder why PSs allow for $R_\Omega\neq 0$ in the  Newtonian stable branch, unlike the scalar stars. Whereas  we do not have a final explanation, we would like  to point out a special feature of PSs. Fig.~\ref{F0} (right) exhibits some physical quantities (main panel) as well as metric functions~\eqref{E24} and Proca potential functions~\eqref{E25} (inset) for the ``special" solution  with $\xi=1$~\eqref{specialone}. One observes, in particular that the energy density $T^t_t$ has a maximum away from the origin. This is a feature that had already been observed for PSs~\cite{brito2016proca,DiGiovanni:2018bvo} and it is more notorious precisely in the Newtonian branch. By contrast,  for the scalar BSs, the maximum of this energy density is always  at the  origin.

\begin{figure}[h]
			 \centering
			\includegraphics[scale=0.30]{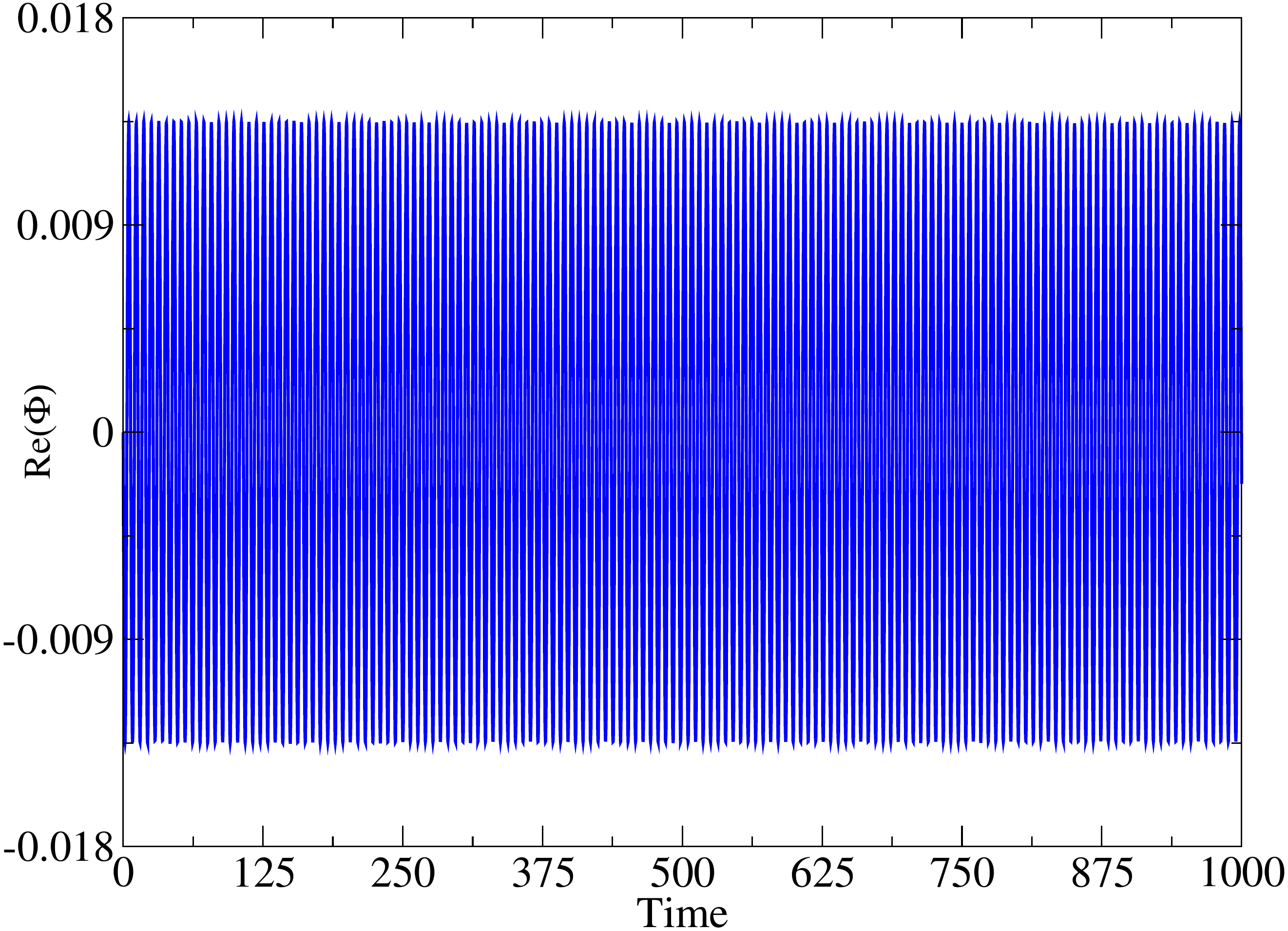}
    \includegraphics[scale=0.30]{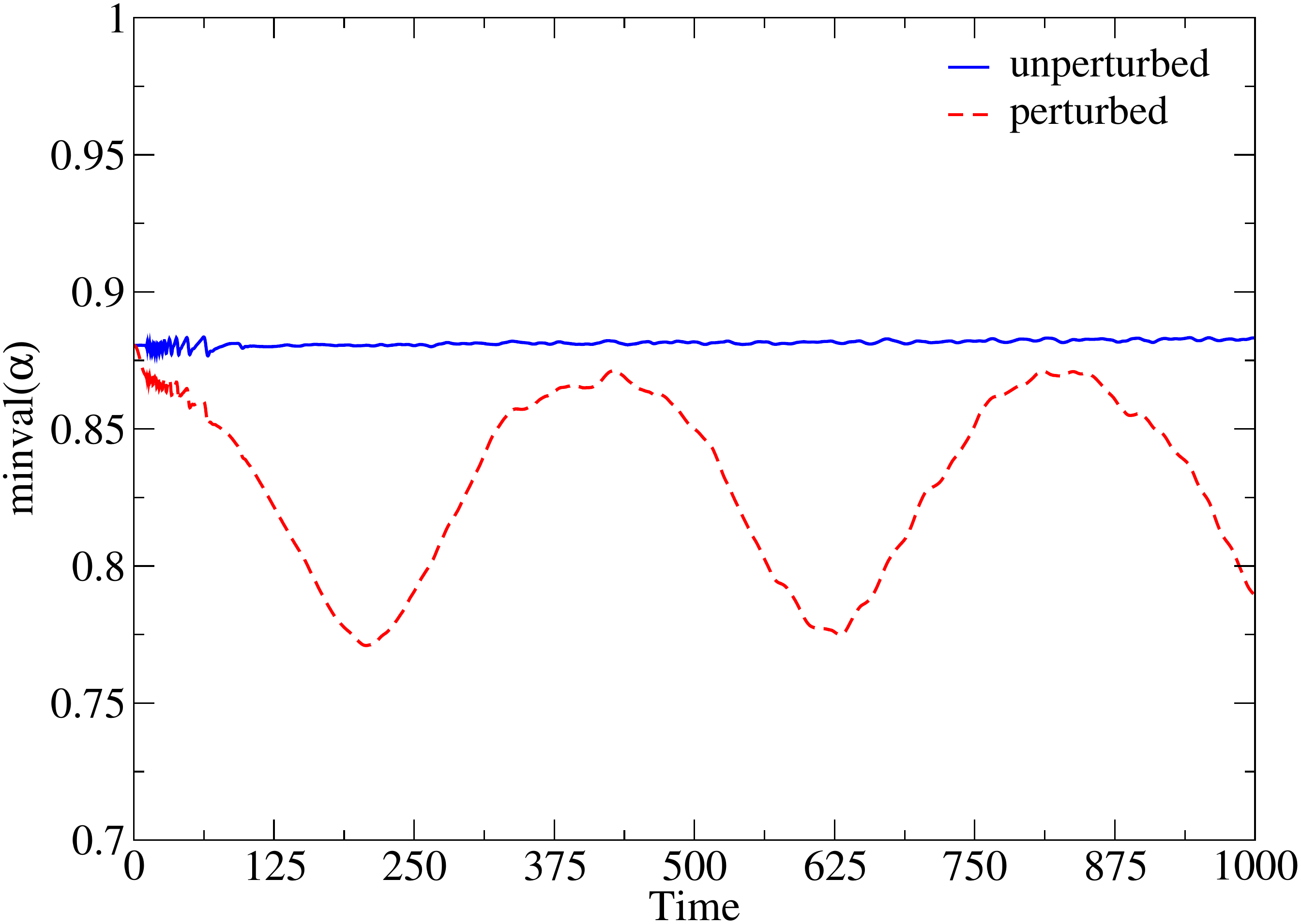}
			\includegraphics[scale=0.3]{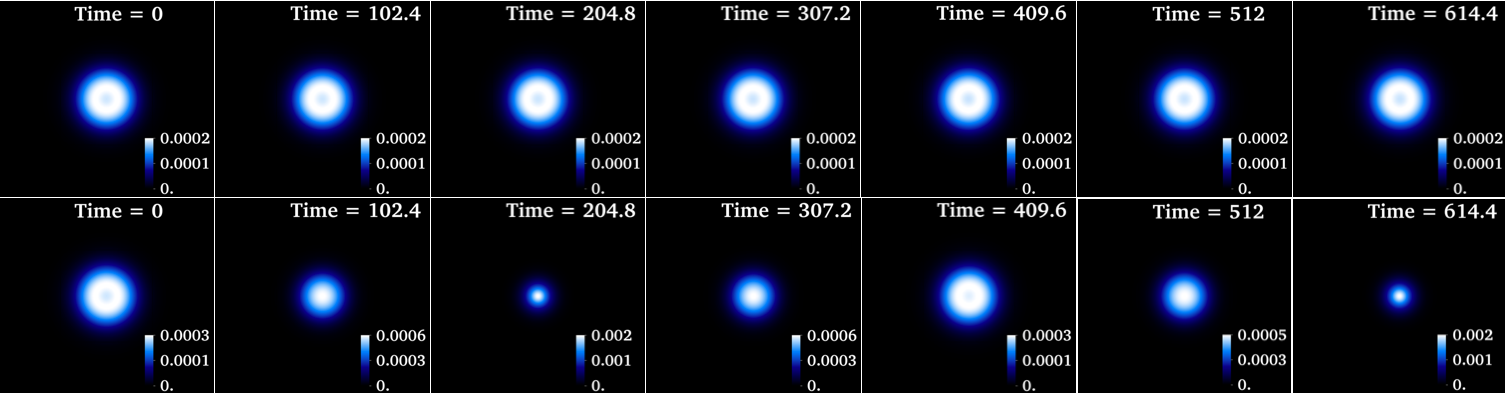}                     
	 		 \caption{Time evolution of  the solution~\eqref{specialone}: (top left) amplitude of the real part of the Proca scalar potential; (top right) minimum value of the lapse [also for a perturbation  of the solution];  (bottom panel) snapshots of the energy density, for both the unperturbed (top row) and perturbed (bottom row) evolutions.}
	 		 \label{F7}
	\end{figure}

To  summarise,  PSs, while not being able to accommodate a LR in the perturbatively stable region, may have  $R_\Omega=R_{\rm ISCO}$ for dynamically stable stars, where the latter refers to a comparable ($i.e.$ same mass), Schwarzschild BH. In order to confirm the dynamical stability of the solution~\eqref{specialone} we have evolved it using similar techniques to the ones described in Section~\ref{S52}. Here we have used the Einstein Toolkit to perform the dynamical evolutions~\cite{EinsteinToolkit,Loffler:2011ay, Zilhao:2013hia}, the Proca equations being solved with a modification of the \textsc{Proca} thorn \cite{Canuda_2020_3565475, Zilhao:2015tya} for a complex Proca field; this setup has been used previously in~\cite{sanchis2017numerical,Sanchis-Gual:2019ljs}. The results are illustrated in Fig.~\ref{F7}. 
The top left panel exhibits the time evolution of the amplitude of the real part of the Proca scalar potential of solution~\eqref{specialone}; this is essentially the  time component of the Proca potential, but see~\cite{sanchis2017numerical} for a precise definition. The maximal amplitude does not change. The top right panel shows the evolution of the minimum value of the lapse; this is essentially the  time-time component of the metric, but see~\cite{sanchis2017numerical} for a precise definition. The two lines refer to the solution~\eqref{specialone} and perturbation of this solution obtained by multiplying the Proca field by  1.05. One observes the unperturbed solution is unaffected by the evolution, whereas the perturbed one oscillates, but does not decay. The bottom panel shows  snapshots of the time evolution of the energy density, for both the unperturbed (top row) and perturbed (bottom row) evolutions. These evolutions clearly confirm the expected stability of the solution~\eqref{specialone}.

To conclude we remark that in the analysis of Proca stars in this Section (and similarly in  the scalar case) we have allowed $\lambda_P$ to take both positive and negative values, as a means to see its impact on  the LR and TCOs, even though  negative  values of $\lambda_P$  lead to a self-interactions potential that is unbounded from below.

\subsection{Lensing}\label{Slen}
Finally, we aim at confirming that the lensing of solution~\eqref{specialone} lit by a thin accretion disk with its inner edge at $R_\Omega=6M$ indeed mimics the shadow of a mass $M$ Schwarzschild BH lit by a similar accretion disk. For this  purpose 
we have used an independent ray-tracing code in order to image the shadow and lensing of both spacetimes. This is the same code that was used in previous works, e.g.~\cite{Cunha:2015yba,Cunha:2016bjh,Cunha:2016bpi,Cunha:2018gql,Cunha:2018cof}, to numerically integrate the null geodesic equations $ \ddot{x}^\mu + \Gamma^\nu_{\alpha \beta} \dot{x}^\alpha \dot{x}^\beta = 0$. This procedure, $i.e.$ backwards ray-tracing, represents the propagation of light rays from the observer backwards in time towards the radiation source or the BH (if it exists).

We consider a very simplified astrophysical setup wherein the only radiation source is an opaque and thin accretion disk located on the equatorial plane around the central compact object. The disk has an inner edge with an areal radius $R_{\Omega}=6M$ in both spacetimes. For the PS this radial location aims to mimic the inner edge of a stalled torus in the equatorial plane, inside which the Magneto-Rotational Instability (MRI) is essentially quenched~\cite{olivares2018tell}. To represent this system, we have imposed a luminosity profile for the disk, with a maximum at the disk edge and with a very fast decay as the radius increases.

The ray-tracing integration of a light ray stops when the photon reaches either: i) the BH, ii) the disk, or iii) numerical infinity. Since the disk is the only light source assumed, photons that never intersect the disk via ray-tracing are shown as black pixels in the image. Black pixels thus include both photons that escape to numerical infinity and that fall into the BH (forming the shadow).

\begin{figure}[t!]
 \centering
\includegraphics[scale=0.20]{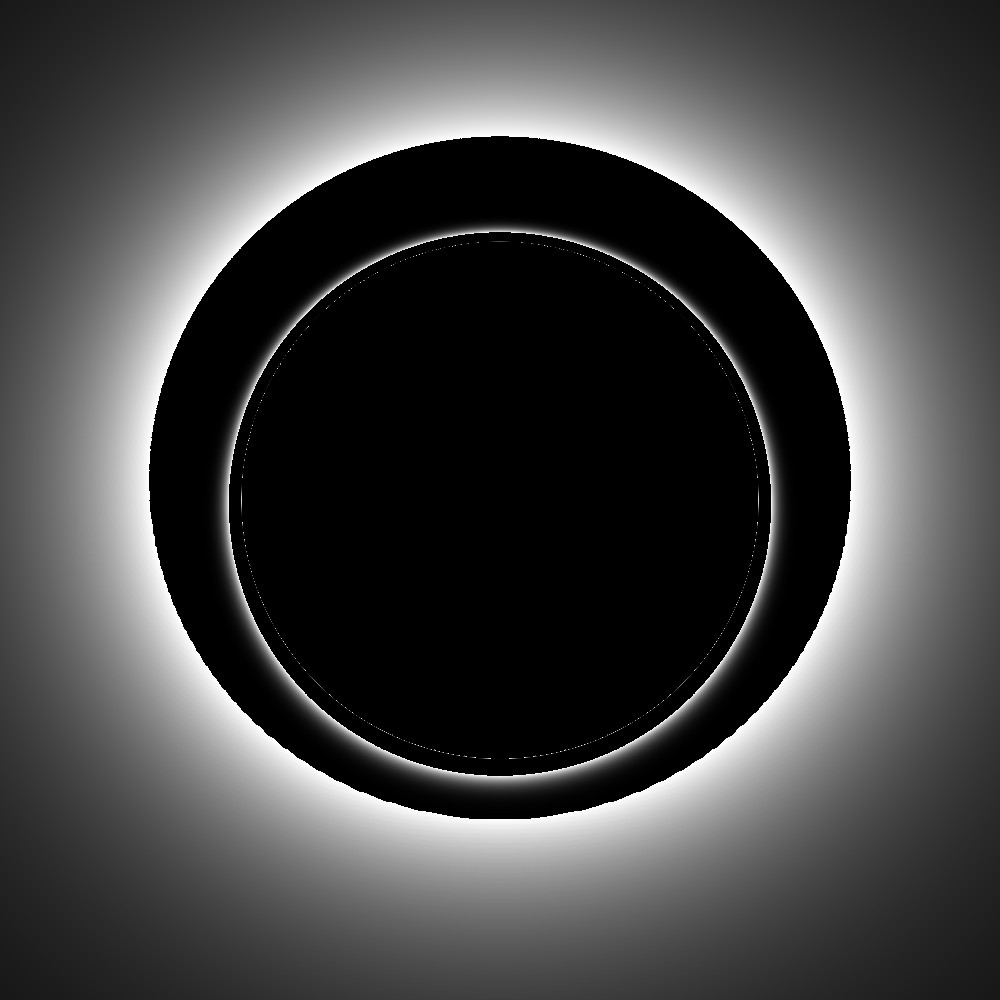}
\includegraphics[scale=0.20]{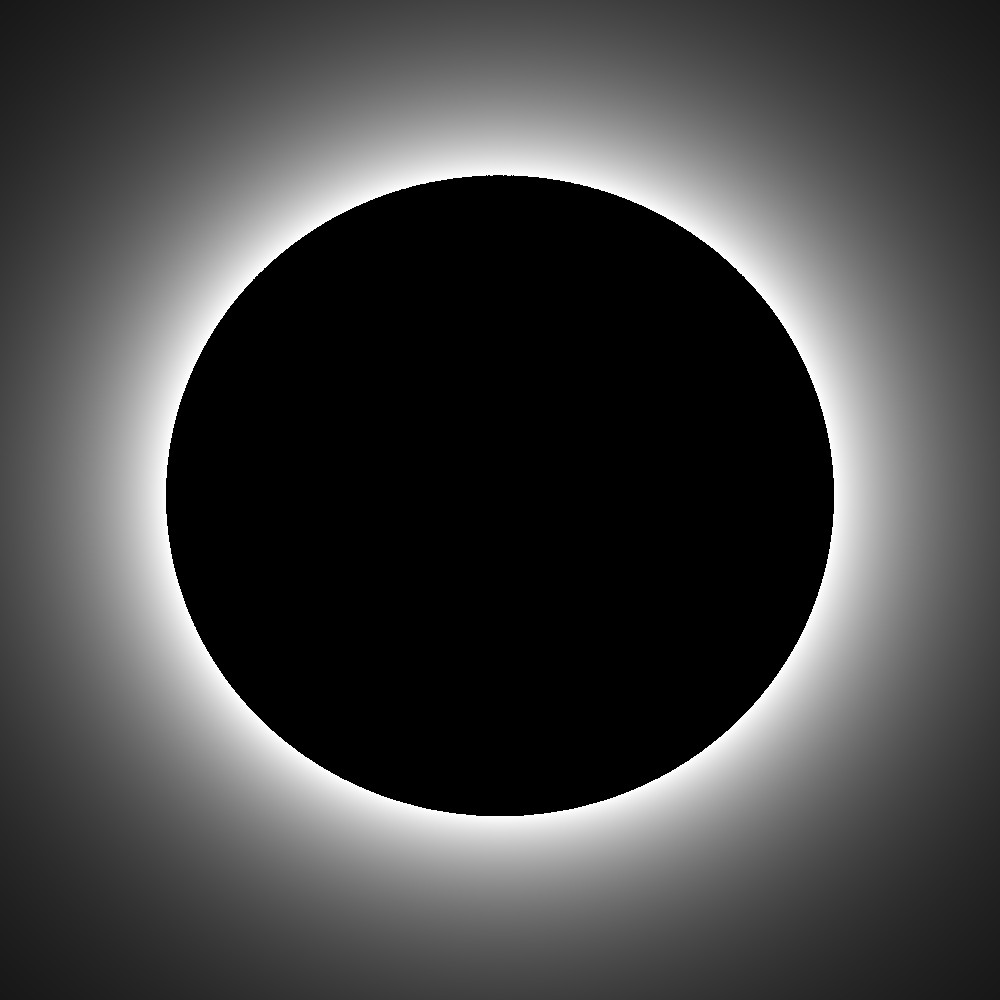}
\includegraphics[scale=0.20]{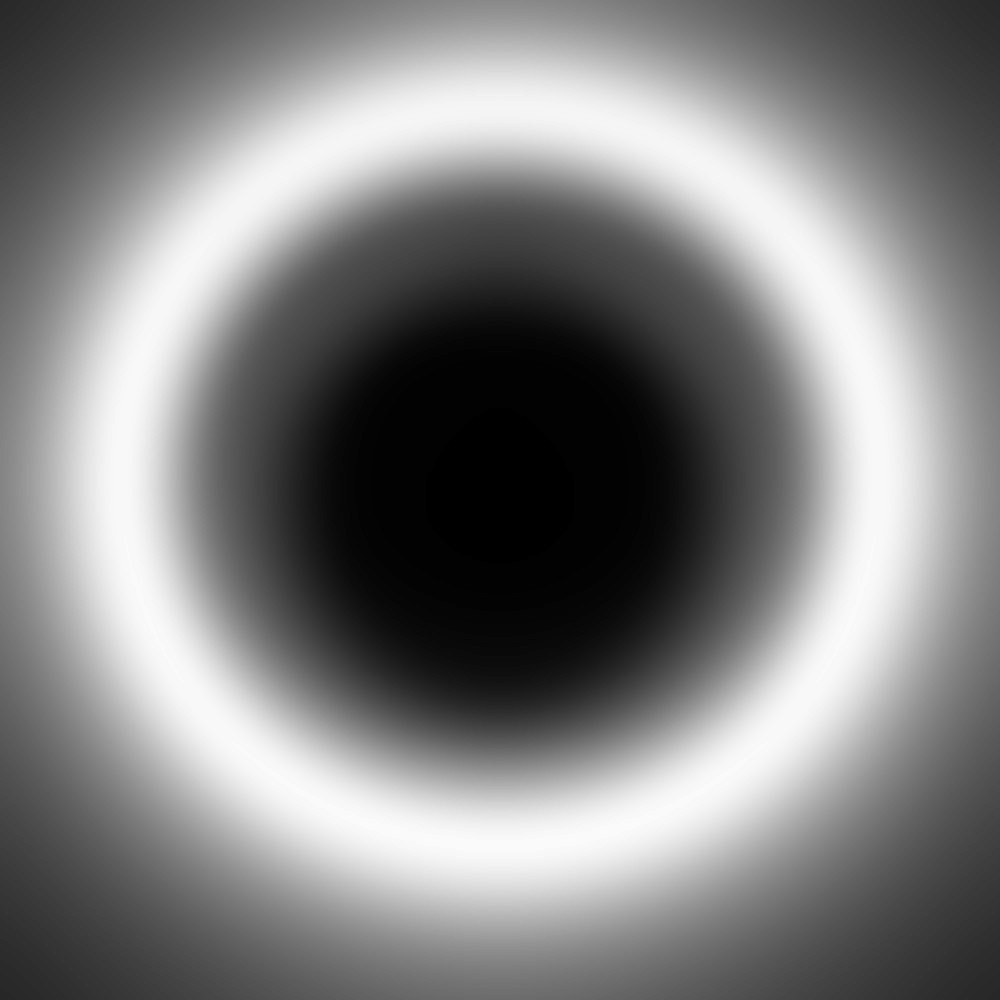}
\includegraphics[scale=0.20]{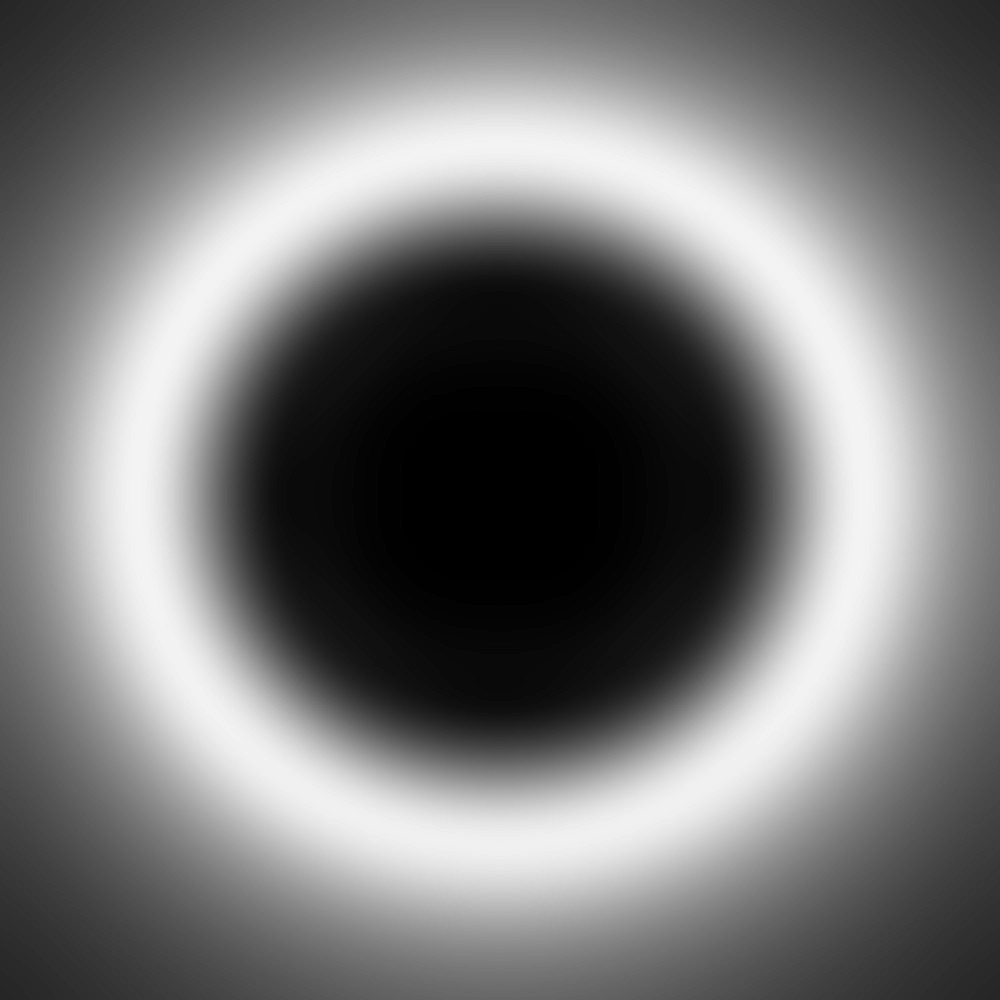}                   
	 		 \caption{Lensing at an observation angle of $17\degree$ (almost polar): (top left) Schwarzschid; (top right) PS; (bottom left) Schwarzschid blurred (bottom right) PS blurred. }
	 		 \label{F8}
	\end{figure}

\begin{figure}[t!]
 \centering
\includegraphics[scale=0.20]{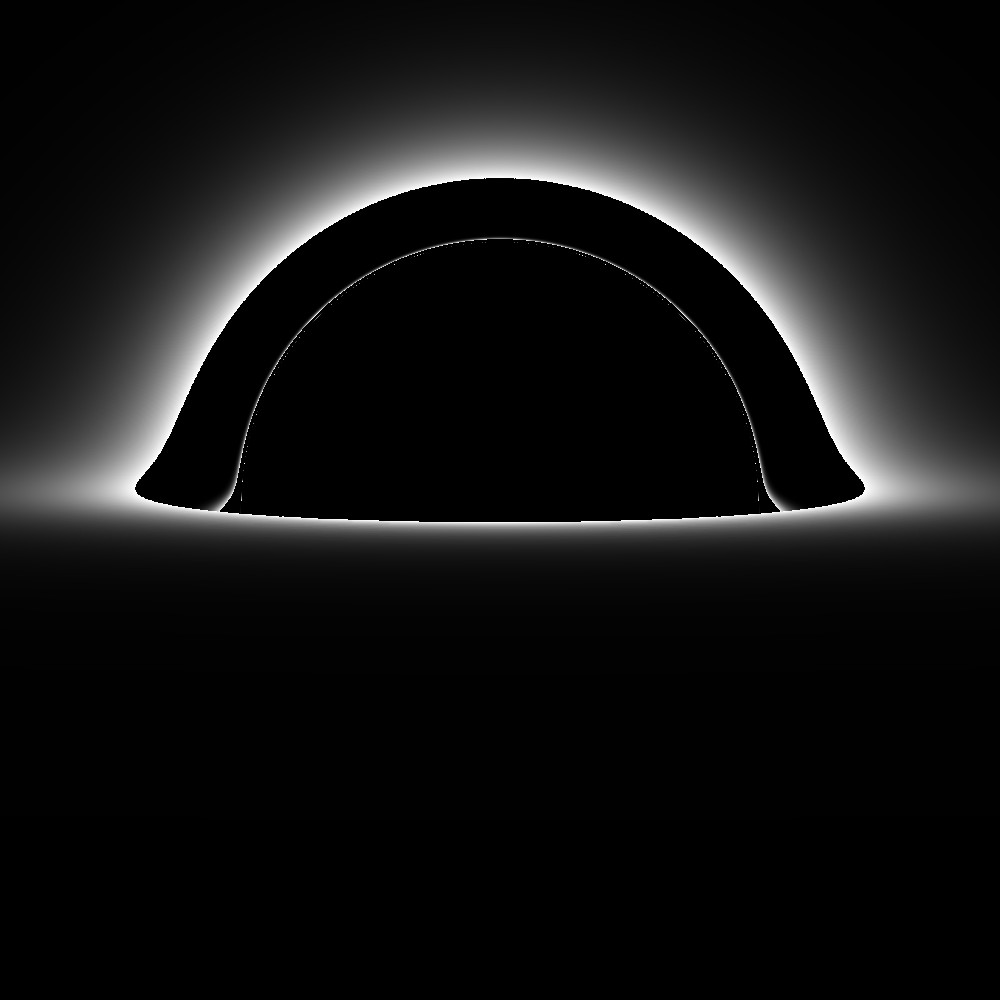}
\includegraphics[scale=0.20]{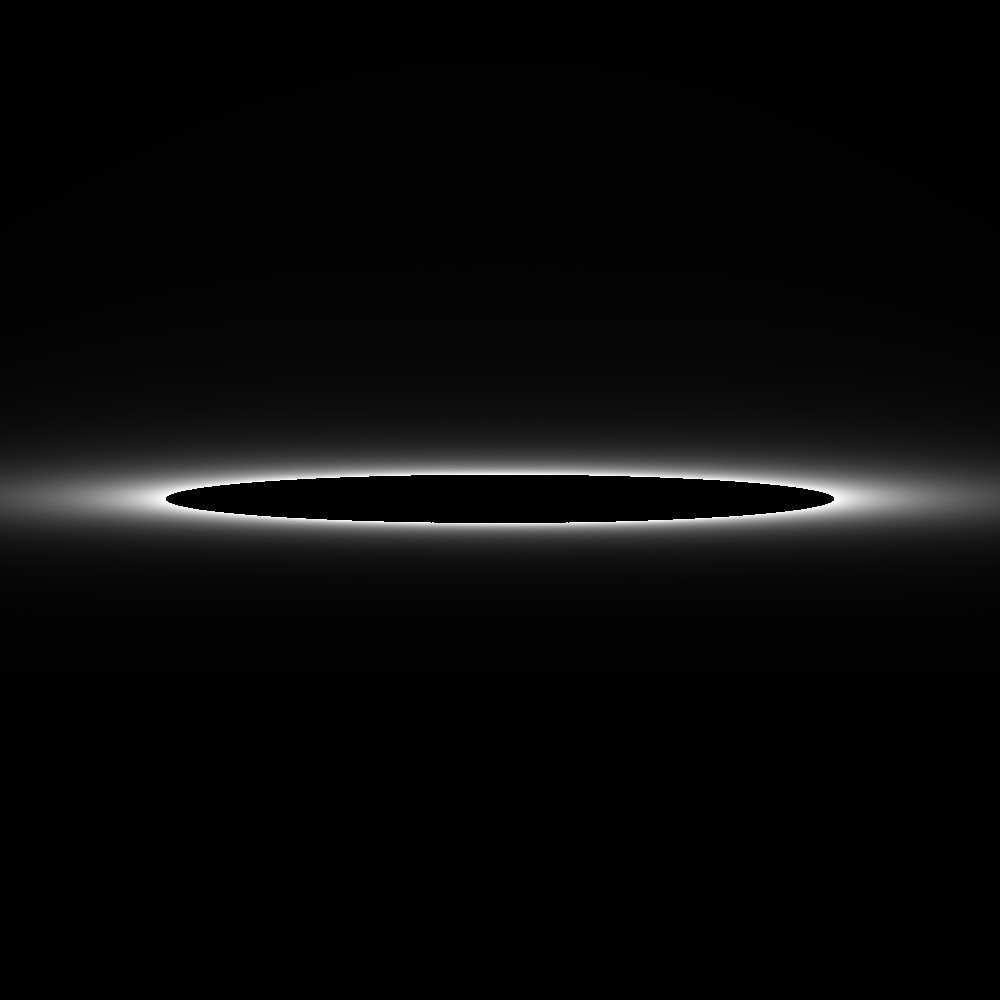}
\includegraphics[scale=0.20]{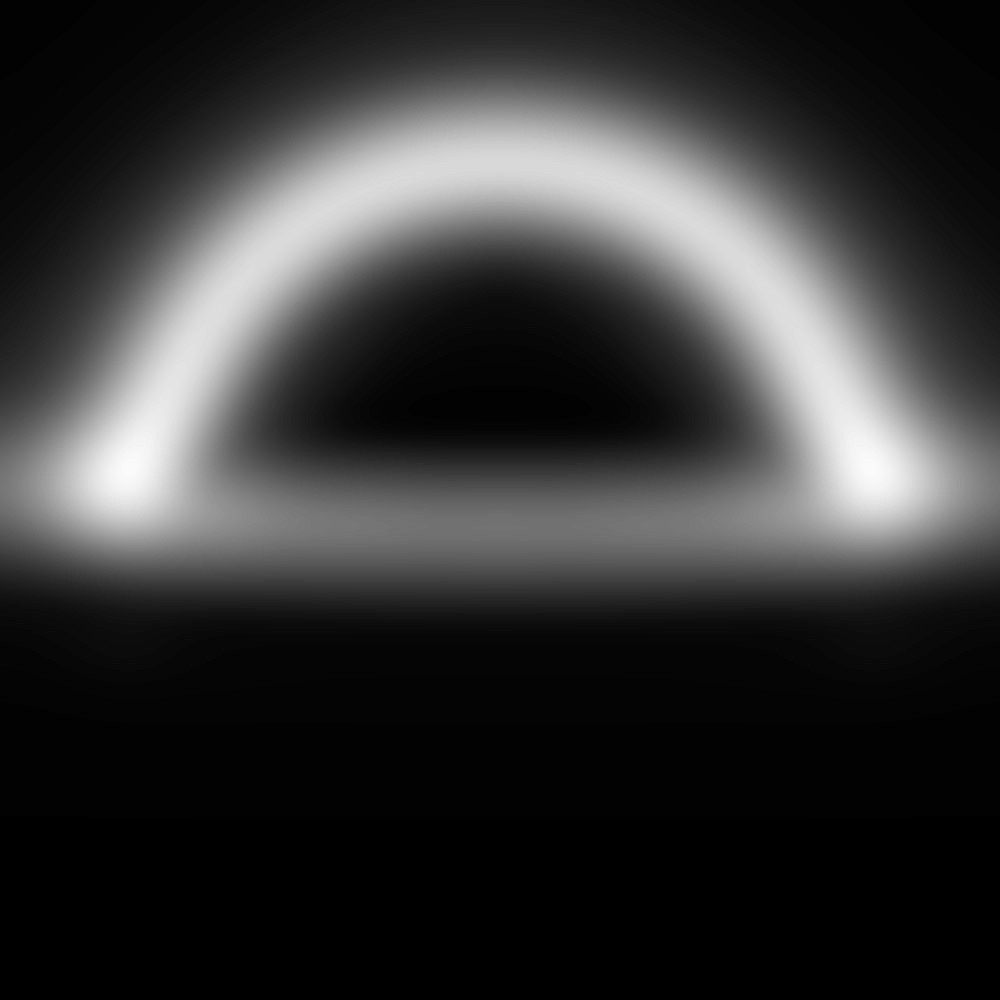}
\includegraphics[scale=0.20]{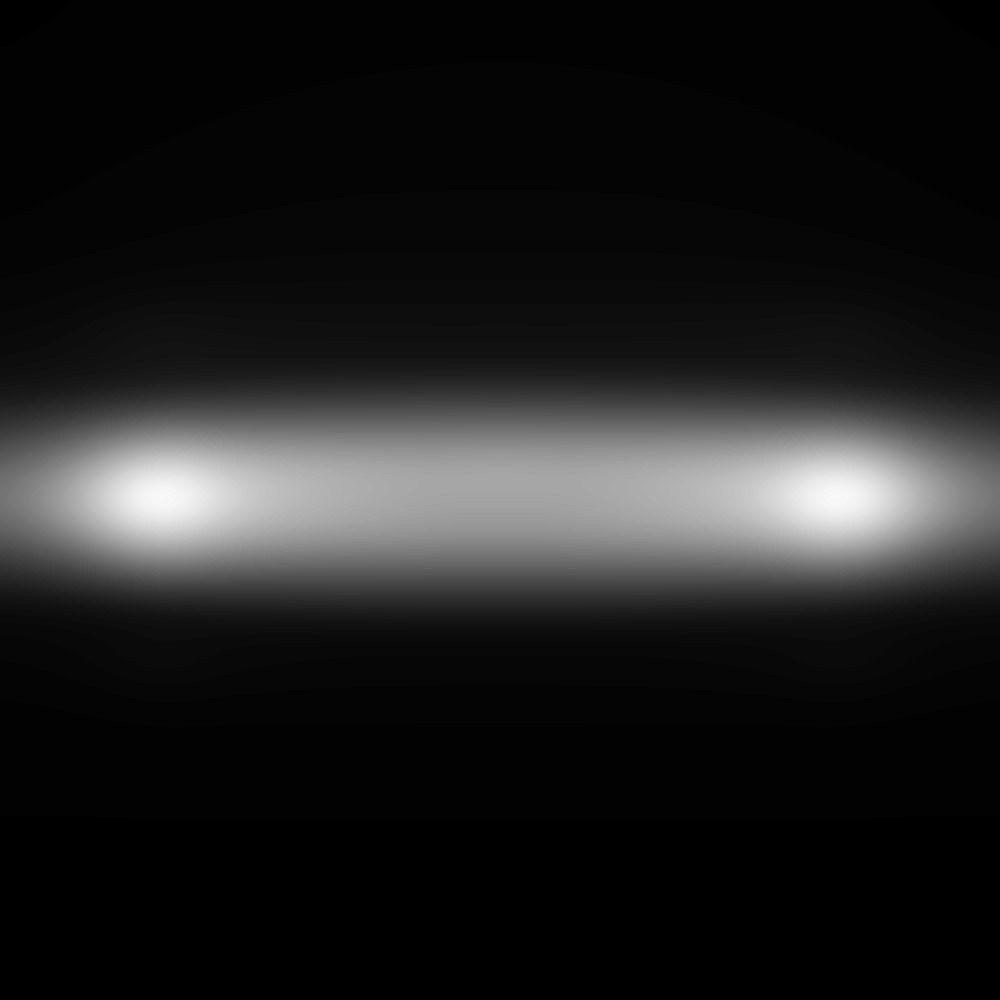}                   
	 		 \caption{Lensing at an observation angle of $86\degree$ (almost equatorial): (top left) Schwarzschid; (top right) PS; (bottom left) Schwarzschid blurred (bottom right) PS blurred.}
	 		 \label{F9}
	\end{figure}

The lensed images are shown  in~Fig.~\ref{F8} and~Fig.~\ref{F9}, and were obtained for an observer placed at an areal radius of $r_o=100M$ with a co-latitude angle $\theta_o=\{17\degree\,,\,86\degree\}$, respectively. Local observation angles were locally discretized into a matrix $1000\times 1000$ of pixels, with both angles varying in the range $\pm\tan^{-1}(1/10)\simeq \pm 5.7º$. This collection of pixels forms the displayed images.

The most interesting case for degeneracy occurs for an observer close to the poles (Fig.~\ref{F8}).  Concretely, the choice $\theta_o=17\degree$, corresponds the angle at which M87*, the target of the Event Horizon Telescope (EHT) 2017 run~\cite{Akiyama:2019cqa}, was observed from Earth. For $\theta_o=17\degree$ the images of the  Schwarzschild  (top left panel) and PS (top right panel) look similar, although some finer additional lensing features are still visible in the Schwarzschild case. Whereas the central dark region in the PS case is due to the lack of source (disk), in the Schwarzschild BH image there is a thin emission ring corresponding to a secondary image of the accretion disk, and indeed higher order images at the very edge of the shadow. Such fine lensing details are absent in the PS "shadow mimicker". The existence of the latter is only a consequence of the assumed absence of light coming from infinity.

The potential similarity between the PS image and Schwarzschild one is further accentuated by considering that current EHT observations have a limited angular resolution of the order of the compact object itself. We can attempt to reproduce this effect by applying a Gaussian blurring filter to the images, which washes away smaller image details. The figures obtained after such a blurring procedure are shown in the bottom panels of Fig.~\ref{F8}, and indeed have an uncanny resemblance with each other, which illustrates how such a PS configuration might potentially mimic a Schwarzschild BH for electromagnetic channel observations.

Let us now analyse a near equatorial observation, choosing $\theta_o=86\degree$. The corresponding images are shown in~Fig.~\ref{F9}. In this case the images of the  Schwarzschild  (top left panel) and PS (top right panel)  are fairly different. In particular, the former resembles the now familiar BH shape displayed in the prominent Hollywood movie {\it Interstellar}~\cite{James:2015yla}, whereas the PS simply looks like what we might have naively expected: an accretion disk with a hole in it, as seen from the side. This is simple to interpret: such PS is  still  fairly Newtonian in some aspects; in particular its gravitational potential well is shallow and so the bending of light it produces is weak. Consequently, the accretion disk has an almost flat spacetime appearence, $i.e.$ a plane with a  hole, whereas in the BH case one sees the background of the disk raised due to considerable light  bending. Again the bottom panels apply the same blurring as in~Fig.~\ref{F8} and  manifest that, even with limited resolution, under this almost equatorial observation, the two objects could be distinguished.

Finally, this discussion aims only to be a proof of concept. The analysis presented herein has several caveats, such as assuming an idealized thin disk without a necessarily physical luminosity profile, as well as not accounting for relativistic effects such as Doppler and gravitational redshifts.  A full GRMHD analysis and ray-tracing is required in the background of this PS to fully settle the question: to what degree can it  imitate a BH observation? The case built herein, nonetheless, clearly confirms  the degeneracy, but only under some observation conditions; under others, the different depth of the potential well impacts decisively in producing a distinguishable image.

%
\section{Discussion}\label{S6}
%
The analysis we have presented in this paper shows that  models in which dynamically robust spherical BSs can have a degenerate (effective) shadow with a comparable Schwarzschild BH \textit{do exist}.

In the case of scalar BSs, we have established  that the most common models of BSs cannot mimic key phenomenological properties of BHs, such as the lensing of light or the accretion flow of matter, if one imposes that these stars should be dynamically stable. Self-interaction cannot easily solve this issue; but as illustrated by the axionic model, appropriate self-interaction terms with sufficiently significant couplings may solve the issue. This is certainly an  interesting possibility,  which could be explored by approximation techniques  for very large self-interactions, like the ones suggested in~\cite{colpi1986boson}, rather than a full numerical approach.

On the other hand, for PSs, we have found that the simplest model, without self-interactions, can indeed mimic a comparable Schwarzschild BH, in the sense of having  the new scale $R_\Omega=6M$, thus equal to the ISCO areal  radius of the BH. As a word of caution, we remark that despite the matching between $R_\Omega$ for the PS and the ISCO of the comparable Schwarzschild BH, the  lensing in the different spacetime geometries leads to a slightly different shadow  size (as a careful inspection of Fig.~\ref{F8} (top panels) reveals).\footnote{This is  in the same spirit  that in a BH spacetime lit from a faraway celestial sphere the shadow size is not determined by the  areal radius of the LR but rather by the impact parameter of the LR photons.} But, since along the Newtonian stable branch of PSs,  $R_\Omega$ varies  from a large  value down to zero, a precise shadow degeneracy will be achieved by a neighbouring solution  of  the special  solution~\ref{specialone}. More importantly, our lensing analysis reveals the degeneracy only holds in certain degeneracy conditions. Interestingly, these include conditions similar to the ones for the M87* observations reported by the Event Horizon Telescope.  It  would be very intereting to perform general relativistic hidrodynamical simulations on these PS backgrounds, similar to the ones in~\cite{olivares2018tell} for the scalar case, to confirm  this degeneracy. 

There are two key assumptions in the conclusions of the last paragraph: the BSs are near equilibrium and are spherical. Firstly, typical stars and BH candidates are spinning; do these results carry through to the spinning case? The answer is  two-fold. Concerning the LRs,  several spinning BSs models have been discussed in the literature, and their LRs have been  computed - see $e.g.$~ \cite{Cunha:2015yba,cunha2016chaotic,grandclement2017light}. In all  cases  LRs emerge beyond  the first mass extremum. In the spinning  case, however,  perturbative stability computations showing that  the mass extremum  coincides with the crossing from stability to instability are absent. Moreover, it has been show that scalar spinning BSs are unstable even in the region naively considered to be stable, whereas spinning PSs appear dynamically robust~\cite{Sanchis-Gual:2019ljs}. Concerning the TCOs, spinning BSs can have an ISCO, unlike the spherical case - see $e.g.$~\cite{Grandclement:2014msa,vincent2016imaging,Franchini:2016yvq,grould2017comparing,Delgado:2020udb}. But it  remains to see  if there is  any model in  which the accretion flow really mimics that of a comparable BH (with the  same mass and angular momentum) and which is,  moreover, dynamically robust.  

Even  if  they  cannot  be ultracompact, stable BSs are nonetheless compact objects that can be evolved in binaries. Recently, an intriguing degeneracy has been established for  GW190521~\cite{Abbott:2020tfl}, showing that a collision of spinning PSs  can fit the observed waveform with a slight statistical preference with respect to the vanilla binary BH  model~\cite{CalderonBustillo:2020srq}.   Thus, even if BSs cannot simply imitate BHs in all of their phenomenology, one cannot exclude that a population of BSs coexists with BHs, as part of the dark matter population,  in particular within a certain mass range, which would be determined by the mass of the ultralight bosonic particle(s). In  this sense, the existence of dynamically robust PSs that can imitate the BH lensing, as shown here, brings to the limelight the issue of degeneracy in lensing/shadow observations.

%
\section*{Acknowledgements}
%
We also thank H. Olivares for   discussions concerning Ref.~\cite{olivares2018tell}. A. Pombo is supported by the FCT grant PD/BD/142842/2018. P. Cunha is supported by the grant BIPD/UI97/7484/2020. This work is supported by the Center for Research and Development
in Mathematics and Applications (CIDMA) through the Portuguese
Foundation for Science and Technology
(FCT - Funda\c c\~ao para a Ci\^encia e a Tecnologia),
references UIDB/04106/2020 and UIDP/04106/2020 and by national funds (OE), through FCT, I.P., in the scope of the framework contract foreseen in the numbers 4, 5 and 6 of the article 23, of the Decree-Law 57/2016, of August 29,
changed by Law 57/2017, of July 19. We acknowledge support  from the projects PTDC/FIS-OUT/28407/2017, CERN/FIS-PAR/0027/2019 and PTDC/FIS-AST/3041/2020
.   This work has further been supported by  the  European  Union's  Horizon  2020  research  and  innovation  (RISE) programme H2020-MSCA-RISE-2017 Grant No.~FunFiCO-777740. The authors would like to acknowledge networking support by the
COST Action CA16104. Computations were performed in the clusters ``Argus'' and ``Blafis'' at the U. Aveiro.


  \bibliographystyle{ieeetr}
  \bibliography{biblio}


\end{document}